\documentclass[twocolumn]{aastex631}
\usepackage{CJK} 

\usepackage{amsmath}
\usepackage{amssymb}
\usepackage{epsfig}
\usepackage{apjfonts}
\usepackage{natbib}
\usepackage{hyperref}
\usepackage{epstopdf}
\usepackage{verbatim}
\usepackage{enumitem}
\usepackage{color}
\setlist[enumerate]{itemsep=-1mm}
\usepackage{soul, xcolor}
\setstcolor{blue}

\usepackage{mathtools}

\usepackage{bigints}

\newcommand{\dotdeg}{\rlap{.}^\circ}
\newcommand{\dotarcsec}{\rlap{.}''}

\submitjournal{AJ (Received: August 3rd, 2022; Accepted: October 12nd, 2022)}

\begin{document}


\begin{CJK*}{UTF8}{gbsn}

\title{The McDonald Accelerating Stars Survey (MASS): Architecture of the Ancient Five-Planet Host System Kepler-444}

\author[0000-0002-3726-4881]{Zhoujian Zhang (张周健)}\thanks{NASA Sagan Fellow}
\affiliation{Department of Astronomy \& Astrophysics, University of California, Santa Cruz, 1156 High St, Santa Cruz, CA 95064, USA}
\affiliation{Department of Astronomy, The University of Texas at Austin, Austin, TX 78712, USA}

\author[0000-0003-2649-2288]{Brendan P. Bowler}
\affiliation{Department of Astronomy, The University of Texas at Austin, Austin, TX 78712, USA}

\author[0000-0001-9823-1445]{Trent J. Dupuy}
\affiliation{Institute for Astronomy, University of Edinburgh, Royal Observatory, Blackford Hill, Edinburgh, EH9 3HJ, UK}

\author[0000-0003-2630-8073]{Timothy D. Brandt}
\affiliation{Department of Physics, University of California, Santa Barbara, Santa Barbara, CA 93106, USA}

\author[0000-0003-0168-3010]{G. Mirek Brandt}
\affiliation{Department of Physics, University of California, Santa Barbara, Santa Barbara, CA 93106, USA}

\author[0000-0001-9662-3496]{William D. Cochran}
\affiliation{Center for Planetary Systems Habitability and McDonald Observatory, The University of Texas at Austin, Austin, TX 78712, USA}

\author[0000-0002-7714-6310]{Michael Endl}
\affiliation{McDonald Observatory and the Department of Astronomy, The University of Texas at Austin, Austin, TX 78712, USA}

\author{Phillip J. MacQueen}
\affiliation{McDonald Observatory and the Department of Astronomy, The University of Texas at Austin, Austin, TX 78712, USA}

\author[0000-0001-5253-1338]{Kaitlin M. Kratter}
\affiliation{Department of Astronomy, University of Arizona, Tucson, AZ 85721, USA}

\author[0000-0002-0531-1073]{Howard T. Isaacson}
\affiliation{Department of Astronomy, University of California, Berkeley, 501 Campbell Hall \#3411, Berkeley, CA 94720, USA}

\author[0000-0003-4557-414X]{Kyle Franson}\thanks{NSF Graduate Research Fellow}
\affiliation{Department of Astronomy, The University of Texas at Austin, Austin, TX 78712, USA}

\author[0000-0001-9811-568X]{Adam L. Kraus}
\affiliation{Department of Astronomy, The University of Texas at Austin, Austin, TX 78712, USA}

\author[0000-0002-4404-0456]{Caroline V. Morley}
\affiliation{Department of Astronomy, The University of Texas at Austin, Austin, TX 78712, USA}

\author[0000-0003-2969-6040]{Yifan Zhou}\thanks{51 Pegasi b Fellow}
\affiliation{Department of Astronomy, The University of Texas at Austin, Austin, TX 78712, USA}

\begin{abstract}
We present the latest and most precise characterization of the architecture for the ancient ($\approx 11$~Gyr) Kepler-444 system, which is composed of a K0 primary star (Kepler-444~A) hosting five transiting planets, and a tight M-type spectroscopic binary (Kepler-444~BC) with an A-BC projected separation of $66$~au. We have measured the system's relative astrometry using the adaptive optics imaging from Keck/NIRC2 and Kepler-444~A's radial velocities from the Hobby Eberly Telescope, and re-analyzed relative radial velocities between BC and A from Keck/HIRES. We also include the Hipparcos-Gaia astrometric acceleration and all published astrometry and radial velocities into an updated orbit analysis of BC's barycenter. These data greatly extend the time baseline of the monitoring and lead to significant updates to BC's barycentric orbit compared to previous work, including a larger semi-major axis ($a = 52.2^{+3.3}_{-2.7}$~au), a smaller eccentricity ($e = 0.55 \pm 0.05$), and a more precise inclination ($i =85\dotdeg4^{+0\dotdeg3}_{-0\dotdeg4}$). We have also derived the first dynamical masses of B and C components. Our results suggest Kepler-444~A's protoplanetary disk was likely truncated by BC to a radius of $\approx 8$~au, which resolves the previously noticed tension between Kepler-444 A's disk mass and planet masses. Kepler-444~BC's barycentric orbit is likely aligned with those of A's five planets, which might be primordial or a consequence of dynamical evolution. The Kepler-444 system demonstrates that compact multi-planet systems residing in hierarchical stellar triples can form at early epochs of the Universe and survive their secular evolution throughout cosmic time.
\end{abstract}

\section{Introduction} 
\label{sec:introduction}

Stellar multiple systems are ubiquitous products of the star formation processes \citep[e.g.,][]{1991A&A...248..485D, 1992ApJ...396..178F, 2010ApJS..190....1R, 2013ARA&A..51..269D, 2022arXiv220310066O}. Thus, a substantial fraction of exoplanets might form in dynamical environments sculpted by stellar multiplicity, with distinct formation histories and orbital architectures from those with single stellar hosts. Close stellar binaries (with the semi-major axes, $a$, below a few au) can possess circumbinary protoplanetary disks massive enough to form P-type planets orbiting both stars \citep[e.g.,][]{2011Sci...333.1602D, 2019ApJ...883...22C}, while wide-separation binaries (with $a$ above a few tens of au) can host S-type planets orbiting either the primary or the secondary star \citep[e.g.,][]{2003ApJ...599.1383H, 2015ApJ...799..170C}. 

The binarity of planet hosting stars is expected to suppress planet formation, as stellar binaries can truncate the protoplanetary disk of either component \citep[e.g.,][]{1994ApJ...421..651A, 2015ApJ...800...96L, 2015MNRAS.452.2396M}; trigger disk turbulence and dynamically excite planetesimals' eccentricities and velocities \citep[e.g.,][]{2006Icar..183..193T, 2015ApJ...798...69R, 2015ApJ...798...71S}; and induce secular oscillations in planets' orbital inclinations and eccentricities via the Kozai--Lidov mechanism \citep[e.g.,][]{1962AJ.....67..591K, 1962P&SS....9..719L, 2013MNRAS.431.2155N}. Indeed, observational studies have shown that the occurrence rate of exoplanets in stellar binaries tends to be smaller than those of wider binaries or single stars \citep[e.g.,][]{2014ApJ...783....4W, 2016AJ....152....8K, 2021MNRAS.507.3593M, 2021AJ....162..192Z}. Moreover, the orbits of planet-hosting stellar binaries appear to be statistically aligned with those of the planets, while orbital inclinations of binaries without planets are likely isotropic \citep[e.g., ][]{2022AJ....163..160B, 2022AJ....163..207C, 2022MNRAS.512..648D}. The orbital alignment between binaries and planets could be primordial if both stellar components and the planets all form within the same massive disk or hierarchical cloud fragmentation that preserves orbital angular momenta \citep[e.g.,][]{2018ApJ...857...40S, 2018AJ....155..160T, 2022AJ....163..207C}. Alternatively, for stellar binary systems formed in misaligned orbits with the protoplanetary disk, the presence of a wide stellar companion can torque the gaseous disk into alignment by inducing disk precession and subsequent energy dissipation \citep[e.g.,][]{2000MNRAS.317..773B, 2012Natur.491..418B, 2018MNRAS.477.5207Z, 2022AJ....163..207C}.

As a hierarchical triple planet-host system, Kepler-444 \citep{2015ApJ...799..170C} provides an excellent laboratory for studying the impact of stellar multiplicity on the formation and dynamical evolution of planetary systems. Located at a distance of $36.52 \pm 0.02$~pc \citep[][]{2021AJ....161..147B}, this system is composed of a K0 dwarf (Kepler-444~A) and a tight \citep[$\lesssim 0.3$~au;][]{2016ApJ...817...80D} M-type spectroscopic binary (Kepler-444~BC) with a projected separation of $1\dotarcsec8$ (or $\approx 66$~au) from A. Kepler-444~A hosts a compact planetary system ($a = 0.04 - 0.08$~au) of five transiting planets with sub-Earth sizes ($R_{p} = 0.4-0.7$~$R_{\oplus}$) and mildly eccentric orbits \citep[$e = 0.1-0.3$;][]{2015ApJ...799..170C, 2019A&A...630A.126B}. Orbital periods of these planets ($3-10$~days) are close to, though not exactly matching, mean-motion resonances \citep{2015ApJ...799..170C}. Due to their proximity to the 5:4 resonance, planets Kepler-444~d and e induce significant transit timing variations in Kepler light curves, leading to measured photodynamical masses of $0.036^{+0.065}_{-0.020}$~M$_{\oplus}$ for the planet d and $0.034^{+0.059}_{-0.019}$~M$_{\oplus}$ for e \citep[][]{2017ApJ...838L..11M}. These two planets thus have low densities, suggestive of water-rich or pure-silicate compositions.

One of the most astounding properties of this complex planetary system is its very old age of $\approx 11$~Gyr, as supported by asteroseismology \citep[e.g.,][]{2015ApJ...799..170C, 2019A&A...630A.126B}, stellar isochrones \citep[e.g.,][]{2016ApJS..225...32B, 2017AJ....154..108J}, a long stellar rotation period \citep[e.g.,][]{2015ApJ...801....3M, 2021NatAs...5..707H}, and the system's Galactic thick-disk membership \citep[e.g.,][]{2015ApJ...799..170C}. Kepler-444~A is metal-poor ([Fe/H] = $-0.52 \pm 0.12$~dex) with enhanced $\alpha$-abundance \citep[][]{2018A&A...612A..46M}, consistent with the observed trends that compact multi-planet systems are more prevalent around metal-poor stars than metal-rich stars \citep[e.g.,][]{2018ApJ...867L...3B} and that metal-poor stars with planets tend to have higher [$\alpha$/Fe] than those without planets \citep[e.g.,][]{2012A&A...547A..36A}. Kepler-444 also belongs to the Arcturus stellar stream \citep[][]{2006A&A...449..533A} which likely has an extragalactic origin \citep[e.g.,][]{2009ApJ...700.1794B, 2014A&A...562A..71B}.

Constraining the barycentric orbit of Kepler-444~BC relative to A provides boundary conditions on the size and mass of the protoplanetary disk that resided around A, informs past and future dynamical interactions between the BC binary and the inner planets, and places this system in the context of statistical studies of planet-hosting stellar binaries \citep[e.g.,][]{2022AJ....163..160B, 2022AJ....163..207C, 2022MNRAS.512..648D}. \cite{2016ApJ...817...80D} provided the first constraints of Kepler-444~BC's barycentric orbit by combining A's multi-epoch radial velocities (RVs), the relative radial velocity between the BC and A components, as well as relative astrometry from three-years of monitoring using adaptive optics (AO) imaging. They found that BC has a highly eccentric orbit ($e \approx 0.86$), leading to a small A--BC separation of $\approx 5$~au at periastron. This implies that the protoplanetary disk of Kepler-444~A was truncated and severely depleted of planet-forming solid material. 

We have acquired new observations of Kepler-444 as part of the McDonald Accelerating Stars Survey \citep[][]{2021AJ....161..106B, 2021ApJ...913L..26B}, an AO imaging program targeting stars with long-term RV trends and astrometric accelerations from Hipparcos and Gaia, which supplement the published astrometric and RV data used in \cite{2016ApJ...817...80D}. Our new relative astrometry of this system extends the time baseline of monitoring to 9~years and our new radial velocities bridge the epochs of two published datasets spanning a total of 12~years. The arrival of high-precision Gaia astrometry \citep[][]{2016A&A...595A...1G}, when combined with Hipparcos, further informs the orbit analysis by providing the sky-projected astrometric acceleration \citep[e.g.,][]{2018ApJS..239...31B, 2021ApJS..254...42B, 2019MNRAS.490.1120F, 2020ApJ...904L..25C, 2021AJ....161..106B, 2021ApJ...913L..26B, 2021AJ....162..266L, 2022MNRAS.513.5588B, 2022AJ....163...50F, 2022arXiv220502729K}, which complements the line-of-sight acceleration revealed by the primary star's radial velocities.

Here we combine our new observations and all published relative astrometry, absolute astrometry, and radial velocities of Kepler-444 to provide the latest constraints on the orbital architecture of this system. Our orbit analysis also sheds new insight into the properties of Kepler-444's protoplanetary disk. We describe our new observations of Kepler-444 in Section~\ref{sec:obs} and the extracted astrometry and radial velocities in Section~\ref{sec:astrometry_rv}. We then present the orbit analysis in Section~\ref{sec:orbit} and discuss their physical implications in Section~\ref{sec:discussion}, followed by a brief summary in Section~\ref{sec:conclusion}.

\section{Observations} 
\label{sec:obs}

\subsection{Adaptive Optics Imaging} 
\label{subsec:keck_ao}

We acquired natural guide star AO images of Kepler-444 on 2019 July 7 UT and 2022 July 12 UT with Keck/NIRC2 in its narrow field of view configuration \citep[][]{2013PASP..125..798W}. On 2019 July 7 UT, we took 10 frames in $J$ band, with an integration of 0.053~sec per coadd and 50 coadds per exposure. On 2022 July 12 UT, we took 10 frames in $H$ band and 9 frames in $K_{S}$ band with 0.018~sec per coadd and 0.053~sec per coadd, respectively (both with 100 coadds per exposure). Kepler-444~A and Kepler-444~BC are widely separated (by $1\dotarcsec8$) in our images and the BC pair is unresolved, as seen from earlier-epoch NIRC2 data \citep[e.g.,][]{2016ApJ...817...80D}, suggesting a tight B-C separation of $\lesssim 0.3$~au (i.e., 1 pixel). In $J$-band images, Kepler-444~A is offset by $\sim 500$~mas from a round partly transparent coronagraph mask with 300~mas in radius (Figure~\ref{fig:sep_pa}). In other words, the closest separation between Kepler-444~A and this mask's edge (i.e., 200~mas) is more than 6 times wider than the circular radius ($30$~mas) adopted to measure A's centroid (see Section~\ref{subsec:relastro}). Given that components of the Kepler-444 system are all outside the coronagraph mask, their relative astrometry should not be impacted by including this mask in the optical path for our $J$-band images, as suggested by \cite{2016AJ....152...28K}. Dome flats and dark frames were taken on the same night as each science dataset. 

We also download all previously published NIRC2 data of Kepler-444 \citep[by][]{2015ApJ...799..170C, 2016ApJ...817...80D, 2022MNRAS.512..648D} from the Keck Observatory Archive\footnote{\url{https://koa.ipac.caltech.edu/cgi-bin/KOA/nph-KOAlogin}}. These data were all taken in pupil-tracking mode and were observed on 2013 August 7 UT (PI: Kraus), 2014 July 28 UT (PI: Kraus), 2014 August 9 UT (PI: Barclay), 2014 November 30 UT (PI: Kraus), 2015 April 11 UT (PI: Liu), 2015 June 22 UT (PI: Mann), 2015 July 21 UT (PI: Kraus), and 2016 June 16 UT (PI: Ireland). We uniformly re-reduce all these published data along with our new observations to avoid any systematics in the relative astrometry caused by different reduction pipelines used in the literature and our work (Section~\ref{sec:astrometry_rv}).

\begin{deluxetable}{lll}
\setlength{\tabcolsep}{20pt} 
\tablecaption{HET/HRS Relative Radial Velocities} \label{tab:hrs_rv} 
\tablehead{ \multicolumn{1}{l}{Epoch} &  \multicolumn{1}{l}{RV$_{\rm A}$} &  \multicolumn{1}{l}{$\sigma_{\rm RV_{A}}$ } \\ 
\multicolumn{1}{l}{(BJD)} &  \multicolumn{1}{l}{(m s$^{-1}$)} &  \multicolumn{1}{l}{(m s$^{-1}$)}} 
\startdata 
2454779.57424  &   $34.21$   &    $3.44$    \\ 
2455020.91240  &   $21.69$   &    $4.93$    \\ 
2455022.91033  &   $25.32$   &    $4.54$    \\ 
2455049.83642  &   $5.39$   &    $5.56$    \\ 
2455139.58038  &   $8.80$   &    $3.41$    \\ 
2455292.93468  &   $0.62$   &    $4.26$    \\ 
2455322.85609  &   $2.78$   &    $4.12$    \\ 
2455525.55168  &   $6.78$   &    $4.26$    \\ 
2455628.99753  &   $4.46$   &    $5.38$    \\ 
2455686.84674  &   $-12.13$   &    $4.86$    \\ 
2455730.74825  &   $-0.39$   &    $4.92$    \\ 
2455837.66975  &   $-1.40$   &    $3.52$    \\ 
2455869.57189  &   $-13.94$   &    $3.61$    \\ 
2456127.66590  &   $-7.60$   &    $5.72$    \\ 
2456194.68123  &   $-1.69$   &    $3.97$    \\ 
2456202.66150  &   $-8.00$   &    $3.68$    \\ 
2456208.64940  &   $-17.84$   &    $3.23$    \\ 
2456224.60794  &   $-14.73$   &    $3.51$    \\ 
2456363.99758  &   $-21.12$   &    $4.46$    \\ 
2456474.95592  &   $-11.12$   &    $5.98$    \\ 
\enddata 
\end{deluxetable}

\subsection{Radial Velocities} 
\label{subsec:relrv}

We obtained precise RV measurements of Kepler-444~A using the High Resolution Spectrograph \citep[HRS;][]{1998SPIE.3355..387T} of the Hobby Eberly Telescope (HET). We used the 316g5936 HRS configuration with a $2''$ diameter optical fiber to obtain a spectral resolving power of $R \approx$60,000. Twenty visits to the target were obtained in queue scheduled mode \citep[][]{2007PASP..119..556S} between 2008~November~09 and 2013~July~01 UT, along with an I$_2$ gas absorption cell which provided the high precision radial velocity metric. A single spectrum of Kepler-444~A without the I$_2$ cell was obtained on 2008~September~30~UT to serve as the stellar spectral template. All HET/HRS spectra were reduced using an automated IRAF script that performs bias subtraction, scattered light removal, and flat-fielding. We also traced the aperture for each echelon spectral order for one-dimensional spectra extraction and calibrated the wavelength solution from the nightly Th-Ar hollow-cathode lamp spectra. Given that HET/HRS did not contain an exposure meter, we estimated the mid-exposure time to be the average of the exposure start and end time. We compute relative RVs of Kepler-444~A from the observed spectra using the auSTRAL code \citep[][]{2000A&A...362..585E} and list them in Table~\ref{tab:hrs_rv}.

\section{Astrometry and Radial Velocity Analysis}
\label{sec:astrometry_rv}

\begin{figure*}[t]
\begin{center}
\includegraphics[height=5.5in]{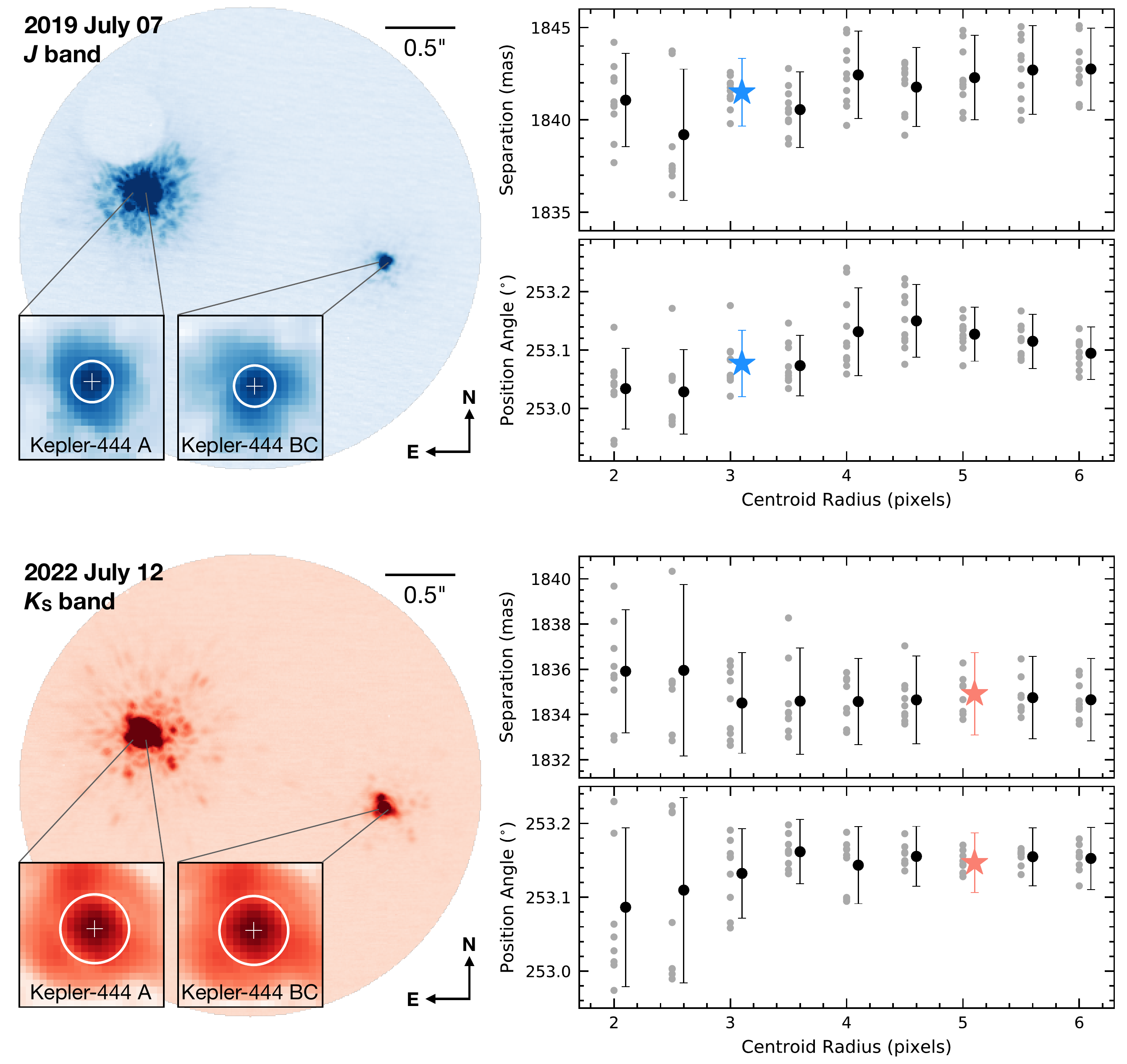}
\caption{{\it Top Left}: A typical reduced and north-aligned $J$-band science frame of Kepler-444 observed on 2019 July 7 UT. Insets present the 20 pixel $\times$ 20 pixel vicinity of A (left) and BC (right) components with their centroids marked by ``+'' signs, computed using a 3 pixel-radius circular region (white circle). A coronagraph mask is visible to the northeast of Kepler-444~A and does not impact our relative astrometry measurements. {\it Top Right}: Centroids of A and BC iteratively computed using a range of circular radii (Section~\ref{subsec:relastro}). At each radius, we show the computed separation and position angle of individual science frames observed on 2019 July 7 UT (grey circle), as well as the resulting separation and position angle measurements with uncertainties computed from Equation~\ref{eq:sep_pa} (black circle). Our final separation and position angle measurements for the $J$-band data are based on a circular radius of 3 pixels and are highlighted as blue stars. {\it Bottom}: Analysis of $K_{S}$-band data observed on 2022 July 12 UT with the same format as the top panel. The white circles in the insets and our final relative astrometry all correspond to a radius of 5 pixels. }
\label{fig:sep_pa}
\end{center}
\end{figure*}

\subsection{Relative Astrometry} 
\label{subsec:relastro}
We (re-)reduce new and published Keck/NIRC2 AO images (Section~\ref{subsec:keck_ao}) in a uniform manner following standard procedures, including applying non-linearity and bad pixel corrections, bias subtraction, flat fielding, and cosmic-ray rejection. The geometric distortions are corrected using the \cite{2010ApJ...725..331Y} solution for data observed before the NIRC2 realignment on 2015 April 13 UT and the \cite{2016PASP..128i5004S} solution for the more recent datasets. We measure the angular separation and position angle of Kepler-444~BC relative to A based on their centroids. For each system component in each distortion-corrected science frame, we first identify the highest-flux pixel and compute a flux-weighted centroid using all data within a certain radius of this brightest pixel. We then iterate this process by updating the circle center to the newly computed centroid position until the relative change in the centroid is below $10^{-6}$. This calculation is carried out for a range of circular radii from 2 to 6 pixels (with intervals of 0.5~pixels) and the final relative astrometry is determined using a radius of $3$~pixels ($30$~mas on the sky) in $J$ band, $4$~pixels ($40$~mas on the sky) in $H$ band, and $5$~pixels ($50$~mas on the sky) in $K'/K_{\rm cont}/K_{S}$ bands as these values correspond to Keck's diffraction limit (Figure~\ref{fig:sep_pa}). 

To evaluate systematic uncertainties of our inferred centroids, we simulate a point spread function (PSF) centered at a random pixel location (fractional pixel locations are allowed) on a detector and then measure its centroid. The PSF is simply described by $I(u) = [2J_{1}(u)/u]^{2}$ with $u = \pi \lambda \theta / D$, where $\theta$ is the angular separation (in units of radians) of a given point on the detector from the PSF center, $J_{1}$ is the Bessel function of its first kind, $D = 10$~m is the aperture diameter of Keck, and $\lambda$ is the effective wavelength of a given NIRC2 filter: $1.2434$~$\mu$m for $J$ band, $1.6197$~$\mu$m for $H$ band, $2.1084$~$\mu$m for $K'$ band, $2.2874$~$\mu$m for $K_{\rm cont}$ band, and $2.1354$ for $K_{S}$ band. We sample the PSF into the pixelated image with two versions of the plate scale as 9.952~mas~pixel$^{-1}$ \citep[][]{2010ApJ...725..331Y} and $9.971$~mas~pixel$^{-1}$ \citep[][]{2016PASP..128i5004S}, corresponding to the detector properties before and after the NIRC2 realignment, respectively. Generating PSFs at random detector locations, we find the differences between the measured and input centroid positions are all below 0.2~mas with a given combination of the band and plate scale. This systematic error is more than $5\times$ smaller than the position uncertainty caused by the distortion correction (see below) and is thus ignored in the error budget of our measured relative astrometry.

\begin{deluxetable*}{lllccc}
\setlength{\tabcolsep}{12pt} 
\tablecaption{Relative Astrometry of Kepler-444} \label{tab:relastro} 
\tablehead{ \multicolumn{1}{l}{Date} &  \multicolumn{1}{l}{Epoch} &  \multicolumn{1}{l}{Filter} &  \multicolumn{1}{c}{Data Reference} &  \multicolumn{1}{c}{Separation} &  \multicolumn{1}{c}{Position Angle} \\ 
\multicolumn{1}{l}{(UT)} &  \multicolumn{1}{l}{(yr)} &  \multicolumn{1}{l}{} &  \multicolumn{1}{l}{} &  \multicolumn{1}{c}{(mas)} &  \multicolumn{1}{c}{($^{\circ}$)}  } 
\startdata 
2013 August 7  &   $2013.598$   &    $K'$   &    \cite{2016ApJ...817...80D}  &   $1842.57 \pm 1.48$  &   $252.911 \pm 0.046$    \\ 
2014 July 28  &   $2014.571$   &    $K_{\rm cont}$   &    \cite{2016ApJ...817...80D}  &   $1843.55 \pm 1.69$  &   $252.876 \pm 0.039$    \\ 
2014 August 9  &   $2014.604$   &    $K'$   &    \cite{2015ApJ...799..170C}  &   $1841.67 \pm 1.62$  &   $252.743 \pm 0.037$    \\ 
2014 November 30  &   $2014.913$   &    $K_{\rm cont}$   &    \cite{2016ApJ...817...80D}  &   $1840.59 \pm 1.61$  &   $252.743 \pm 0.036$    \\ 
2015 April 11\tablenotemark{\scriptsize a}  &   $2015.276$   &    $K_{\rm cont}$   &    \cite{2016ApJ...817...80D}  &   $1840.33 \pm 2.55$  &   $252.760 \pm 0.048$    \\ 
2015 April 11\tablenotemark{\scriptsize a}  &   $2015.276$   &    $K_{\rm cont}$   &    \cite{2016ApJ...817...80D}  &   $1841.41 \pm 1.55$  &   $252.764 \pm 0.034$    \\ 
2015 June 22  &   $2015.473$   &    $K_{\rm cont}$   &    \cite{2022MNRAS.512..648D}  &   $1842.39 \pm 1.75$  &   $252.785 \pm 0.039$    \\ 
2015 July 21  &   $2015.552$   &    $K_{\rm cont}$   &    \cite{2022MNRAS.512..648D}  &   $1841.92 \pm 1.76$  &   $252.783 \pm 0.039$    \\ 
2016 June 16  &   $2016.458$   &    $K_{\rm cont}$   &    \cite{2022MNRAS.512..648D}  &   $1840.78 \pm 1.72$  &   $252.775 \pm 0.047$    \\ 
2019 July 7  &   $2019.514$   &    $J$   &    This Work  &   $1841.50 \pm 1.83$  &   $253.077 \pm 0.057$    \\ 
2022 July 12  &   $2022.527$   &    $H$   &    This Work  &   $1835.78 \pm 1.76$  &   $253.137 \pm 0.045$    \\ 
2022 July 12  &   $2022.527$   &    $K_{S}$   &    This Work  &   $1834.91 \pm 1.82$  &   $253.147 \pm 0.040$    \\ 
\enddata 
\tablenotetext{a}{We distinguish two sets of NIRC2 data taken with different detector sizes and rotator positions following \cite{2016ApJ...817...80D}.}  
\end{deluxetable*}

Given the centroids of BC ($x_{i,{\rm BC}}, y_{i,{\rm BC}}$) and A ($x_{i, {\rm A}}, y_{i, {\rm A}}$) in each science frame (denoted by $i$), the on-detector separation ($r_{i}$; in units of pixels) and position angle ($p_{i}$; in units of degrees) are calculated as:
\begin{equation}
\begin{aligned}
r_{i} &= \big[ (x_{i, \rm BC} - x_{i, \rm A})^{2} + (y_{i, \rm BC} - y_{i, \rm A})^{2} \big]^{1/2 }\\
p_{i} &= {\rm mod}\left[ {\rm - 2 \times arctan}\left(\frac{x_{i, \rm BC} - x_{i, \rm A}}{r_{i} + y_{i, \rm BC} - y_{i, \rm A}} \right)\times 180^{\circ}/\pi, 360^{\circ} \right] 
\end{aligned}
\end{equation}
Here $p_{i}$ becomes $180^{\circ}$ when $x_{i, \rm BC} - x_{i, \rm A} = 0$ and $y_{i, \rm BC} - y_{i, \rm A} < 0$. At a given epoch, we compute these parameters' mean and standard deviation ($\bar{r}$, $\sigma_{r}$; $\bar{p}$, $\sigma_{p}$) over all science frames and convert them into an on-sky separation ($\rho$; in units of mas) and position angle ($\theta$; in units of degree) as \citep[also see Section~4.3 of][]{2018AJ....155..159B}:
\begin{equation} \label{eq:sep_pa}
\begin{aligned}
\rho &= s\bar{r} \\
\sigma_{\rm \rho} &= \rho \big[ (\sigma_{s} / s)^{2} + (\sigma_{r} / \bar{r})^{2} + 2 (\sigma_{d,r} / \bar{r})^{2} \big]^{1/2} \\
\theta &= \bar{p} + \textsc{parang} + \textsc{rotposn} - \textsc{instangl} - \theta_{\rm north} \\
\sigma_{\theta} &= \big[ \sigma_{p}^{2} + \sigma_{\theta,\rm north}^{2} + \left(s \sigma_{d,r} / \rho \times 180^{\circ}/\pi\right)^{2} \big]^{1/2} \\
\end{aligned}
\vspace{-4mm}
\end{equation}
For data taken before (and after) the NIRC2 realignment, we adopt a plate scale $s$ and uncertainty $\sigma_{s}$ as $9.952 \pm 0.002$~mas~pixel$^{-1}$ ($9.971 \pm 0.004$~mas~pixel$^{-1}$), and the north orientation offset $\theta_{\rm north}$ and its uncertainty $\sigma_{\theta,\rm north}$ as $0\dotdeg252 \pm 0\dotdeg009$ ($0\dotdeg262 \pm 0\dotdeg020$) \citep[][]{2010ApJ...725..331Y, 2016PASP..128i5004S}. Here $\sigma_{d,r} = 0.1$~pixels, representing the typical pixel position uncertainty near each component's centroid due to the distortion correction. We extract values of $\textsc{parang}$ (parallactic angle), $\textsc{rotposn}$ (rotator user position), and $\textsc{instangl}$ (zero point of the NIRC2 position angle) from FITS headers of our data. The uniformly measured relative astrometry is summarized in Table~\ref{tab:relastro}. 

Our latest-epoch AO images reveal that the separation and position angle of BC's barycenter relative to A is significantly decreasing and increasing with time, respectively, due to the orbital motion. These trends were not well-constrained based on the astrometric monitoring prior to the year 2017 \citep[e.g.,][]{2016ApJ...817...80D, 2022MNRAS.512..648D}.

\begin{figure}[t]
\includegraphics[height=4.2in]{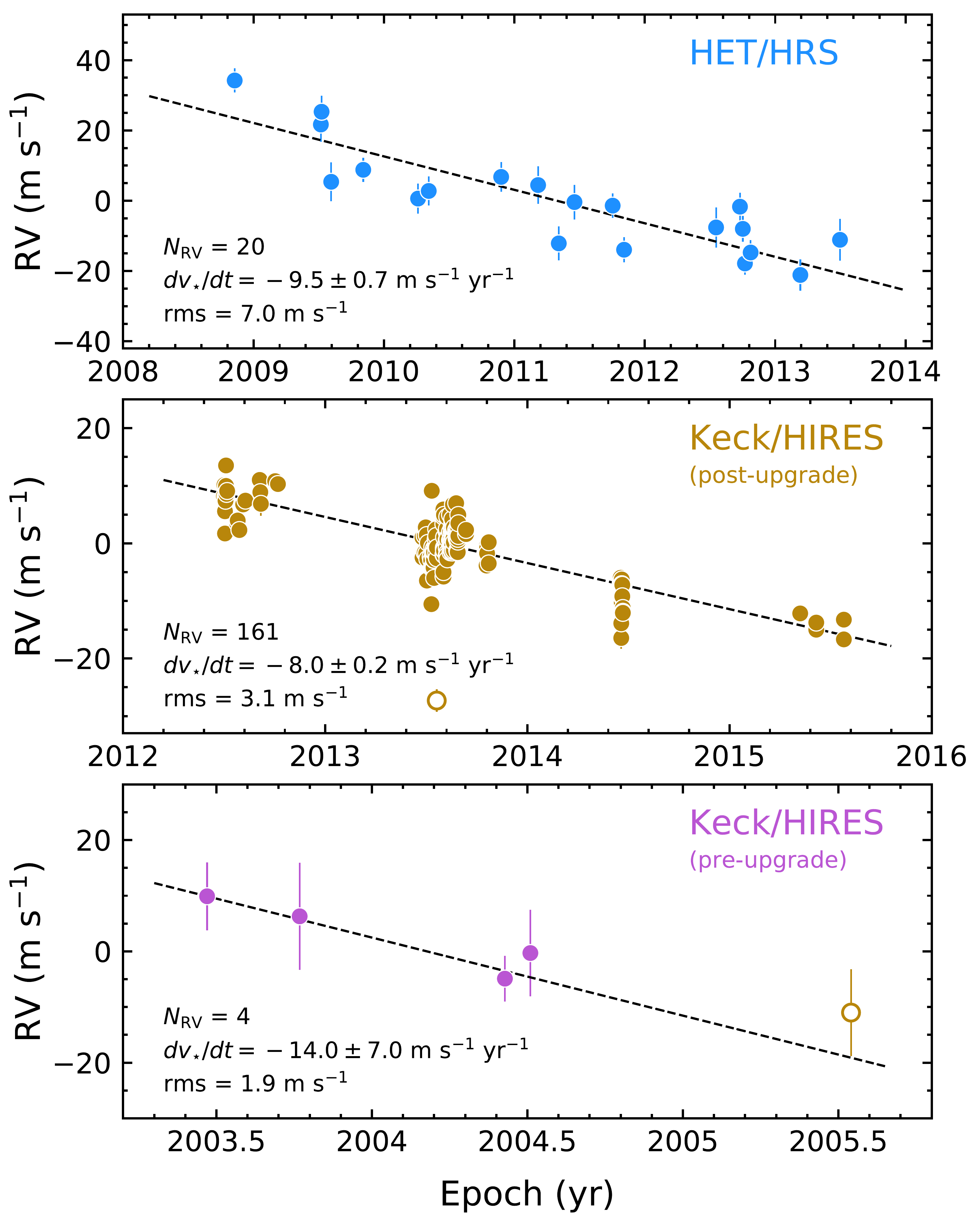}
\caption{Multi-epoch relative RVs of Kepler-444~A measured from HET/HRS (top) in this work and from Keck/HIRES after (middle) and before (bottom) the CCD upgrade on 2004 August 18 UT by \cite{2009ApJ...697..544S}, \cite{2016ApJ...817...80D}, and \cite{2017AJ....153..208B}. We use open circles to mark two relative RV measurements excluded from our analysis (see Section~\ref{subsec:rvtrends}). Linear fits of relative RVs are shown as dashed lines, and we label the fitted RV slopes and rms, as well as the total number of RV measurements used in our orbit analysis.}
\label{fig:relative_rv}
\end{figure}

\subsection{Absolute Astrometry} 
\label{subsec:absastro_hgca}
While Kepler-444~BC was not detected by Hipparcos or Gaia~DR1 at the time of the previous analysis \citep[][]{2016ApJ...817...80D}, both A and BC now have Gaia~EDR3 proper motions of $(\mu_{\alpha}\cos{\delta}, \mu_{\delta}) = (94.64 \pm 0.02, -632.27 \pm 0.02)$~mas~yr$^{-1}$ and $(94.51 \pm 0.05, -630.78 \pm 0.08)$~mas~yr$^{-1}$, respectively \citep[][]{2016A&A...595A...1G, 2021A&A...649A...1G}, which is particularly useful for constraining the BC-to-A mass ratio \citep[e.g.,][]{2021AJ....162..186B}. Also, Kepler-444~A exhibits a significant difference between its Gaia and the joint Hipparcos-Gaia long-term proper motions \citep[reduced $\chi_{\nu}^{2} = 1052$ for a constant proper-motion model;][]{2021ApJS..254...42B}, equivalent to an astrometric acceleration of $20.8 \pm 0.5$~m~s$^{-1}$~yr$^{-1}$.

\subsection{RV Acceleration of Kepler-444~A} 
\label{subsec:rvtrends}
Our HET/HRS RVs of Kepler-444~A show a significant linear trend of $-9.5 \pm 0.7$~m~s$^{-1}$~yr$^{-1}$ with an rms of 7.0~m~s$^{-1}$ (Figure~\ref{fig:relative_rv}). We also collect published RVs of Kepler-444~A from Keck/HIRES, including 163 epochs of RVs measured after the HIRES CCD upgrade on 2004 August 18 UT \citep[][]{2009ApJ...697..544S, 2016ApJ...817...80D, 2017AJ....153..208B} and 4 epochs before the upgrade \citep[][]{2009ApJ...697..544S}. We treat these two sets of RV measurements as separate instruments. Among all post-upgrade HIRES RVs, we exclude one relative RV ($-11.0 \pm 7.8$~m~s$^{-1}$) observed on 2005~July~17 UT \citep[][]{2009ApJ...697..544S}. This relative RV measurement lines up with the trend established by the pre-upgrade RVs, but is $9\sigma$ lower than the extrapolated value ($64.3 \pm 1.5$~m~s$^{-1}$) from the RV measurements over 2012--2016, suggesting the measurement made in 2005 has a different RV zero point. We also exclude one relative RV ($-27.30 \pm 1.98$~m~s$^{-1}$) observed on 2013 July 21 UT \citep[][]{2017AJ....153..208B}, which is $14\sigma$ lower than the other relative RVs measured within 2 years (Figure~\ref{fig:relative_rv}). The remaining 161 post-upgrade HIRES RVs exhibit a slope of $-8.0 \pm 0.2$~m~s$^{-1}$~yr$^{-1}$ (rms $=3.1$~m~s$^{-1}$). The four pre-upgrade HIRES RV measurements show a linear trend of $-14.0 \pm 7.0$~m~s$^{-1}$~yr$^{-1}$ (rms $=1.9$~m~s$^{-1}$). The combined HRS and HIRES data comprise 185 RVs together, spanning a baseline of 12 years.

\subsection{Relative RV between BC and A} 
\label{subsec:deltarv}

We perform a re-analysis of the Keck~I/HIRES spectra of Kepler-444~BC that were used by \cite{2016ApJ...817...80D} to measure absolute radial velocities of both B and C components. Multi-epoch absolute RVs of the individual binary components can constrain the systemic RV of this binary. Comparing the absolute RV of the Kepler-444~BC system to that of Kepler-444~A, \cite{2016ApJ...817...80D} measured the orbital speed orthogonal to the plane of the sky and used this in their orbit analysis. Our re-analysis was originally motivated by a discrepancy in our own orbital analysis and that of \cite{2016ApJ...817...80D} with the sign and possibly the amplitude of the BC$-$A relative RV, i.e., $\Delta{{\rm RV}_{BC-A}} = {\rm RV}_{BC} - {\rm RV}_{A}$. We also include one additional HIRES spectrum of Kepler-444~BC, so our re-analysis uses a total of four epochs of BC's RV measurements.

All spectra were obtained in the standard setup of the California Planet Search \citep[CPS;][]{2010ApJ...721.1467H}, which provides consistent wavelength solutions for the three chips.\footnote{\url{https://exoplanets.caltech.edu/cps/hires/}} To define RV zero points, we use the HIRES spectrum of the RV standard Barnard's star \citep[$-110.11$~km\,s$^{-1}$;][]{2018MNRAS.475.1960F}, its barycentric correction of $-22.75$~km\,s$^{-1}$, and the barycentric corrections of Kepler-444~BC over the four epochs of $-5.39$~km\,s$^{-1}$, $-7.04$~km\,s$^{-1}$, $4.38$~km\,s$^{-1}$, $-0.02$~km\,s$^{-1}$, respectively. For each spectral order, we interpolate the science spectrum and the standard spectrum onto a common wavelength grid, which is uniform in $\log{(\lambda)}$ and has the same number of pixels as the input spectra. We then use the cross-correlation procedure \textsc{c\_correlate} in IDL to compute the wavelength differences in pixels between Kepler-444~BC and the standard. To convert this pixel shift into radial velocity, we use the median pixel size of $1.29-1.31$~km\,s$^{-1}$\,pix$^{-1}$. We fit the cross-correlation functions as the sum of two Gaussians, each with its own position, amplitude, and standard deviation, plus a linearly sloped background. The best-fit model is derived using the Levenberg-Marquardt algorithm implemented in IDL by the \textsc{mpfit} routine for IDL \citep{2009ASPC..411..251M}. Given that not all HIRES orders provide well-defined double-peaked cross-correlation functions, we only use the best five orders from the red chip in our analysis (orders 1, 3, 4, 5, and 8). Finally, we determine RV$_{\rm B}$ as the position of the higher Gaussian peak and RV$_{\rm C}$ as that of the lower peak. Table~\ref{tab:bc_absrv} summarizes our resulting RVs, where we quote the means and, for error bars, the standard deviations across the different HIRES orders.

Comparing our newly derived absolute RVs to those reported in \citet{2016ApJ...817...80D}, we find excellent agreement in the RV differences between B and C, but the zero points are slightly different by $\approx$1--2\,km\,s$^{-1}$. We believe this is most likely due to small systematic errors (1--2\%) in the pixel scale used in the previous analysis because the zero point offset is the largest at the epochs where the difference in pixels between the standard star and science target is also the largest. 

\begin{figure}[t]
\includegraphics[height=2.8in]{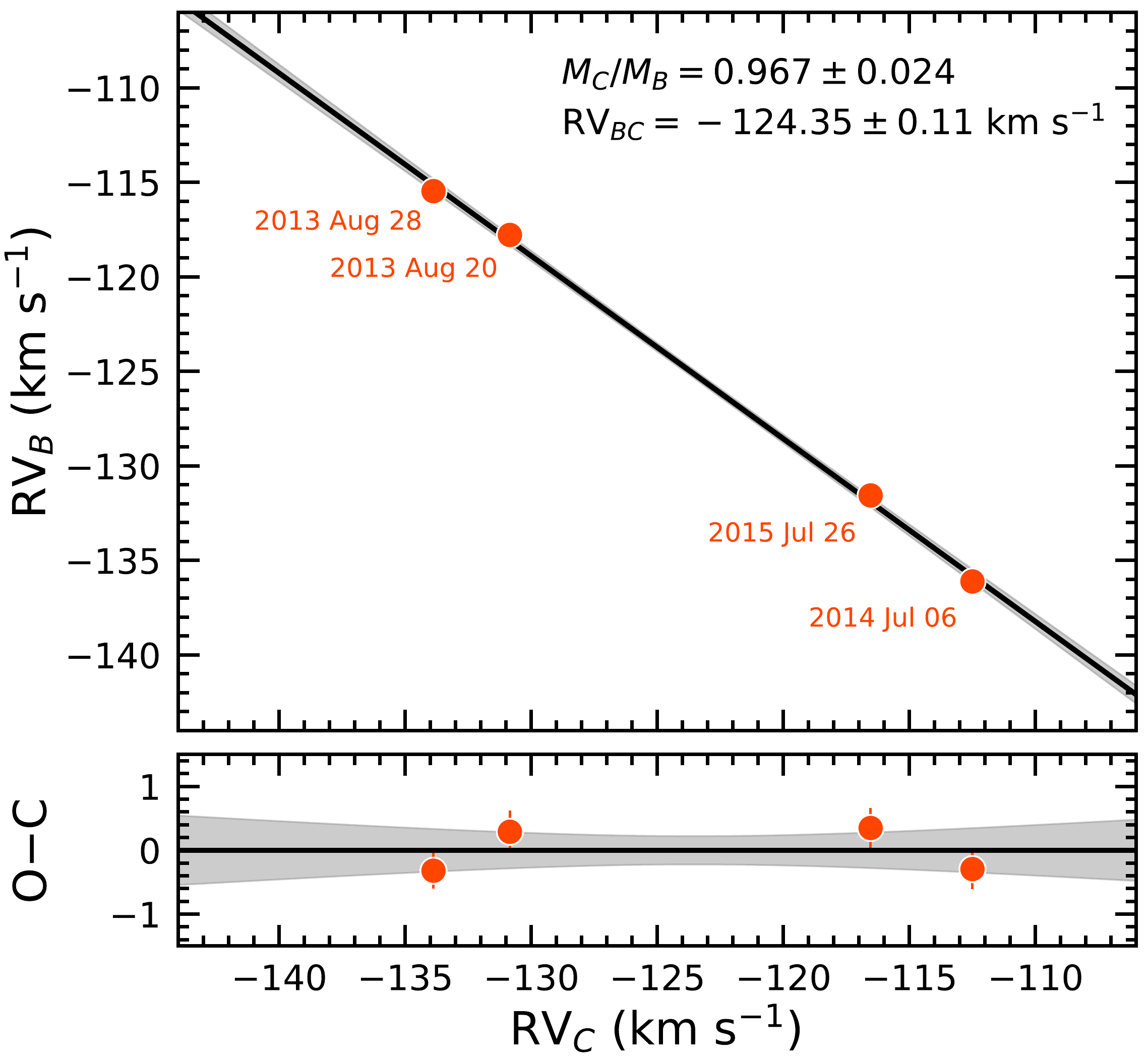}
\caption{{\it Top}: Absolute radial velocities of Kepler-444~B and C components (orange circles; Table~\ref{tab:bc_absrv}), overlaid with the fitted model (black) and the $1\sigma$ interval (grey) as described in Equation~\ref{eq:rv_bc_model}. {\it Bottom}: The observed$-$calculated (i.e., O$-$C) residuals. }
\label{fig:rv_bc}
\end{figure}

Following \cite{1941ApJ....93...29W}, we convert BC's multi-epoch RVs into the systemic velocity RV$_{BC}$ and the C-to-B mass ratio $q_{C-B}$ based on this expression:
\begin{equation} \label{eq:rv_bc_model}
{\rm RV}_{B}  = -q_{C-B} \times {\rm RV}_{C} + {\rm RV}_{BC} \times (1 + q_{C-B})
\end{equation}
We perform an orthogonal distance regression to incorporate the RV uncertainties of each component (Figure~\ref{fig:rv_bc}) and derive RV$_{BC} = -124.35 \pm 0.11$~km~s$^{-1}$, leading to a BC$-$A relative RV of $\Delta{{\rm RV}_{BC-A}} = {\rm RV}_{BC}  - {\rm RV}_{A} = -3.1 \pm 0.2$~km~s$^{-1}$ during the HIRES observations that span 1.9~years. Given that $\Delta{RV_{BC-A}}$ is periodically changing within a full barycentric orbit of Kepler-444~BC, we estimate its time derivative based on A's RV acceleration as:
\begin{equation} \label{eq:derivation_deltarv}
\begin{aligned}
\frac{d}{dt}\big( \Delta{{\rm RV}_{BC-A}} \big) &= \frac{d}{dt}\big({\rm RV}_{BC} - {\rm RV}_{A}\big) \\
&= \frac{d}{dt}\big(- \frac{M_{A}}{M_{BC}} {\rm RV}_{A}\ - {\rm RV}_{A}\big) \\
&= -\frac{M_{A} + M_{BC}}{M_{BC}} \times \frac{d}{dt}\big( {\rm RV}_{A} \big)
\end{aligned}
\end{equation}
The absolute values of Kepler-444~A's RV acceleration are below $20$~m~s$^{-1}$~yr$^{-1}$ (Section~\ref{subsec:rvtrends}). By assuming a very conservative BC-to-A mass ratio\footnote{Our orbit analysis has determined the dynamical mass of Kepler-444~BC with a BC-to-A mass ratio of $0.81 \pm 0.04$ (Section~\ref{sec:orbit} and Table~\ref{tab:orbparams}). Also, \cite{2016ApJ...817...80D} derived a ratio of $0.71 \pm 0.07$ by comparing Kepler-444~BC's photometry-based mass and the Kepler-444~A's asteroseismic mass.} of $0.6$, we estimate that $\Delta{{\rm RV}_{BC-A}}$ increases by $<0.1$~km~s$^{-1}$ over the 1.9-year HIRES observations and this change is smaller than the measured $\Delta{RV_{BC-A}}$ uncertainty. Therefore, we adopt a mean epoch of 2456783.1~JD for this BC$-$A relative RV and include this single-epoch measurement into our subsequent orbit analysis. We have also determined the C-to-B mass ratio as $q_{C-B} = 0.967 \pm 0.024$ (Figure~\ref{fig:rv_bc}), leading to the first individual dynamical masses for B and C components (see Section~\ref{sec:orbit}).

\begin{deluxetable}{ccc}
\setlength{\tabcolsep}{10pt} 
\tablecaption{Absolute Radial Velocities of Kepler-444~BC} \label{tab:bc_absrv} 
\tablehead{ \multicolumn{1}{l}{Epoch} &  \multicolumn{1}{l}{RV$_{B}$} &  \multicolumn{1}{l}{RV$_{C}$} \\ 
\multicolumn{1}{l}{(JD)} &  \multicolumn{1}{l}{(km s$^{-1}$)} &  \multicolumn{1}{l}{(km s$^{-1}$)} } 
\startdata 
2456524.75034  &  $-117.78 \pm 0.11$  &  $-130.85 \pm 0.09$ \\ 
2456532.74431  &  $-115.46 \pm 0.08$  &  $-133.88 \pm 0.12$ \\ 
2456844.98015  &  $-136.11 \pm 0.10$  &  $-112.50 \pm 0.09$ \\ 
2457229.93446  &  $-131.56 \pm 0.10$  &  $-116.54 \pm 0.19$ \\ 
\enddata 
\end{deluxetable}

\begin{figure*}[t]
\begin{center}
\includegraphics[height=6.5in]{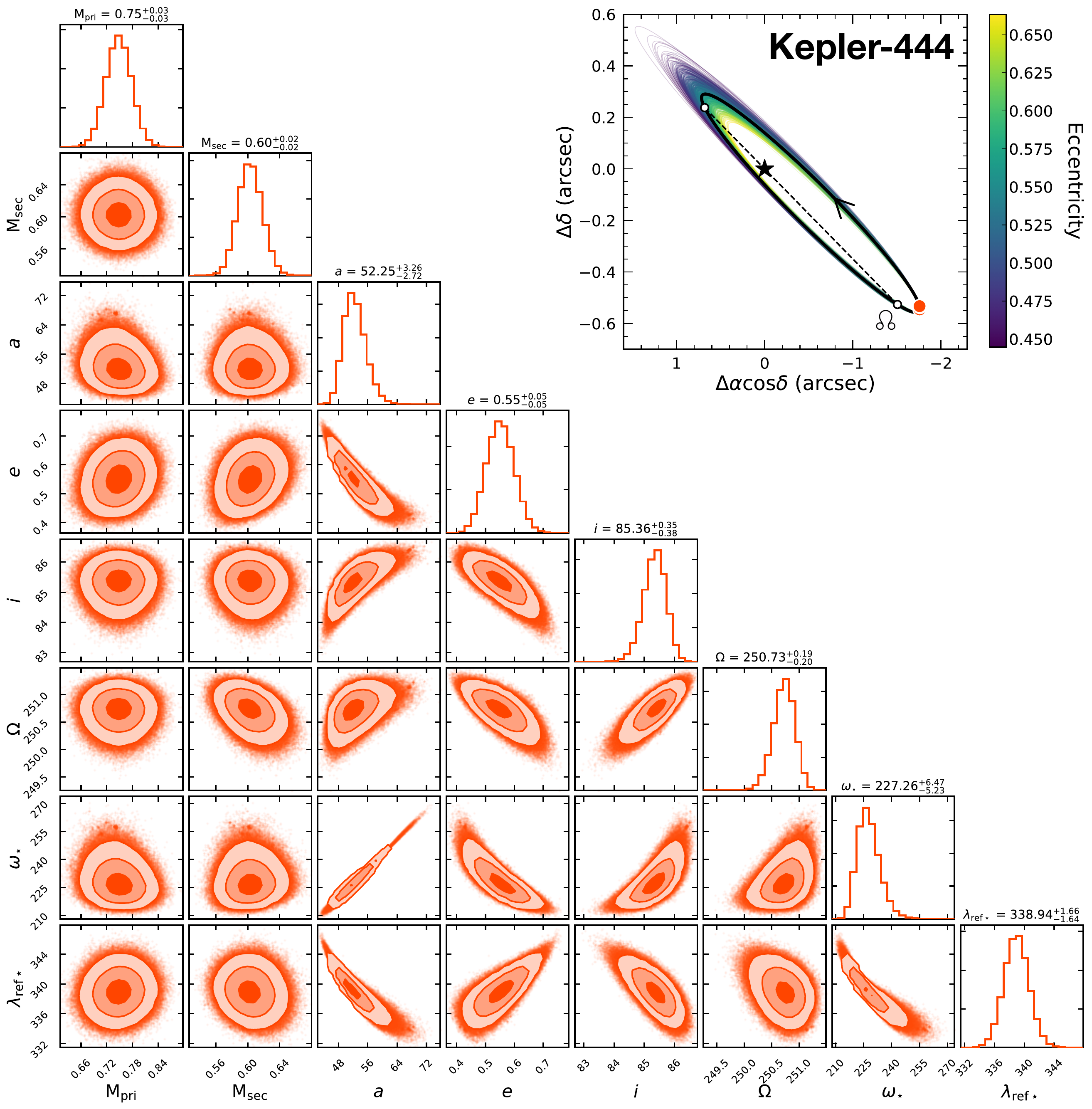}
\caption{Posteriors from our orbit analysis of Kepler-444. Details about each parameter, including credible intervals and the best-fit values of these parameters are listed in Table~\ref{tab:orbparams}. The top right panel shows the predicted relative astrometry between the A and BC components based on 1000 randomly drawn orbits from the MCMC chains, here color-coded by eccentricity. The black solid line shows the best-fit orbit. The two white circles mark the ascending node (i.e., the point in BC's orbit in which it is moving toward the observer through the sky plane; labeled) and the descending node connected via a dashed line (i.e., the line of nodes). Kepler-444~A is shown as a black star and the observed relative astrometry of BC traces out the orbital arc at the bottom right.  }
\label{fig:orvara_corner}
\end{center}
\end{figure*}

\section{Orbit Analysis}
\label{sec:orbit}
We use {\tt orvara} \citep[][]{2021AJ....162..186B} to constrain the barycentric orbit and the dynamical mass of Kepler-444~BC by combining the system's relative astrometry, Hipparcos-Gaia absolute astrometry, Kepler-444~A's multi-epoch RVs, and the single-epoch BC-A relative RV (Section~\ref{sec:astrometry_rv}). We use the parallel-tempering Markov Chain Monte Carlo (MCMC) sampler \citep[][]{2013PASP..125..306F, 2016MNRAS.455.1919V} and run 50 temperatures and 100 walkers over $10^{6}$ steps (per walker) to fit for 17 free parameters, including the masses of Kepler-444~A ($M_{A}$) and BC ($M_{BC}$), semi-major axis of the system ($a$), eccentricity ($e$), inclination ($i$), argument of the periastron of the primary star's orbit ($\omega_{\star}$), position angle of the ascending node ($\Omega$), mean longitude of the primary star's orbit at epoch J2010.0 (i.e., $2455197.5$~JD; $\lambda_{\rm ref,\star}$), the marginalized parallax ($\varpi$) and barycentric proper motion ($\mu_{\alpha}\cos{\delta}$ and $\mu_{\delta}$) of the system, and three combinations of the RV jitter ($\sigma_{\rm jit}$) and RV zero point (ZP) for the HET/HRS, pre-upgrade Keck/HIRES, and post-upgrade Keck/HIRES datasets.\footnote{All orbital parameters correspond to the secondary's orbit unless noted. Also, $e$ and $\omega_{\star}$ are implicitly fitted as $\sqrt{e}\sin{\omega_{\star}}$ and $\sqrt{e}\cos{\omega_{\star}}$ following the convention of {\tt orvara}.} We save the chains every 50 steps and remove the first 5000 samples from each walker of the thinned chains as burn-in.

We set a Gaussian prior for the primary star's mass as $M_{A} = 0.75 \pm 0.03$~M$_{\odot}$, which is derived by \cite{2019A&A...630A.126B} using the same stellar oscillation frequencies but the updated stellar spectrophotometric properties and different sets of evolution models from those in \cite{2015ApJ...799..170C}. This derived mass is also consistent with those in previous studies \citep[e.g.,][]{2015ApJ...799..170C, 2018A&A...612A..46M, 2019A&A...622A.130B}. In Appendix~\ref{app:orbit_broadprior}, we demonstrate that our fitted orbital parameters remain nearly unchanged if we adopt a broader prior on the mass of Kepler-444~A as $0.75 \pm 0.15$~M$_{\odot}$. Log-flat priors are used for $M_{BC}$, $a$, and $\sigma_{\rm jit}$ (constrained between $10^{-5}$~m~s$^{-1}$ and $10$~m~s$^{-1}$), and an isotropic distribution prior is assumed for $i$. Uniform priors are used for $\sqrt{e}\sin{\omega_{\star}}$, $\sqrt{e}\cos{\omega_{\star}}$, $\Omega$, $\lambda_{\rm \star, ref}$, $\mu_{\alpha}\cos{\delta}$, $\mu_{\delta}$, and RV ZPs. A Gaussian prior is set for $\varpi$ with the mean and standard deviation from Gaia~EDR3 ($27.358 \pm 0.013$~mas). 

\begin{figure*}[t]
\begin{center}
\includegraphics[height=8.5in]{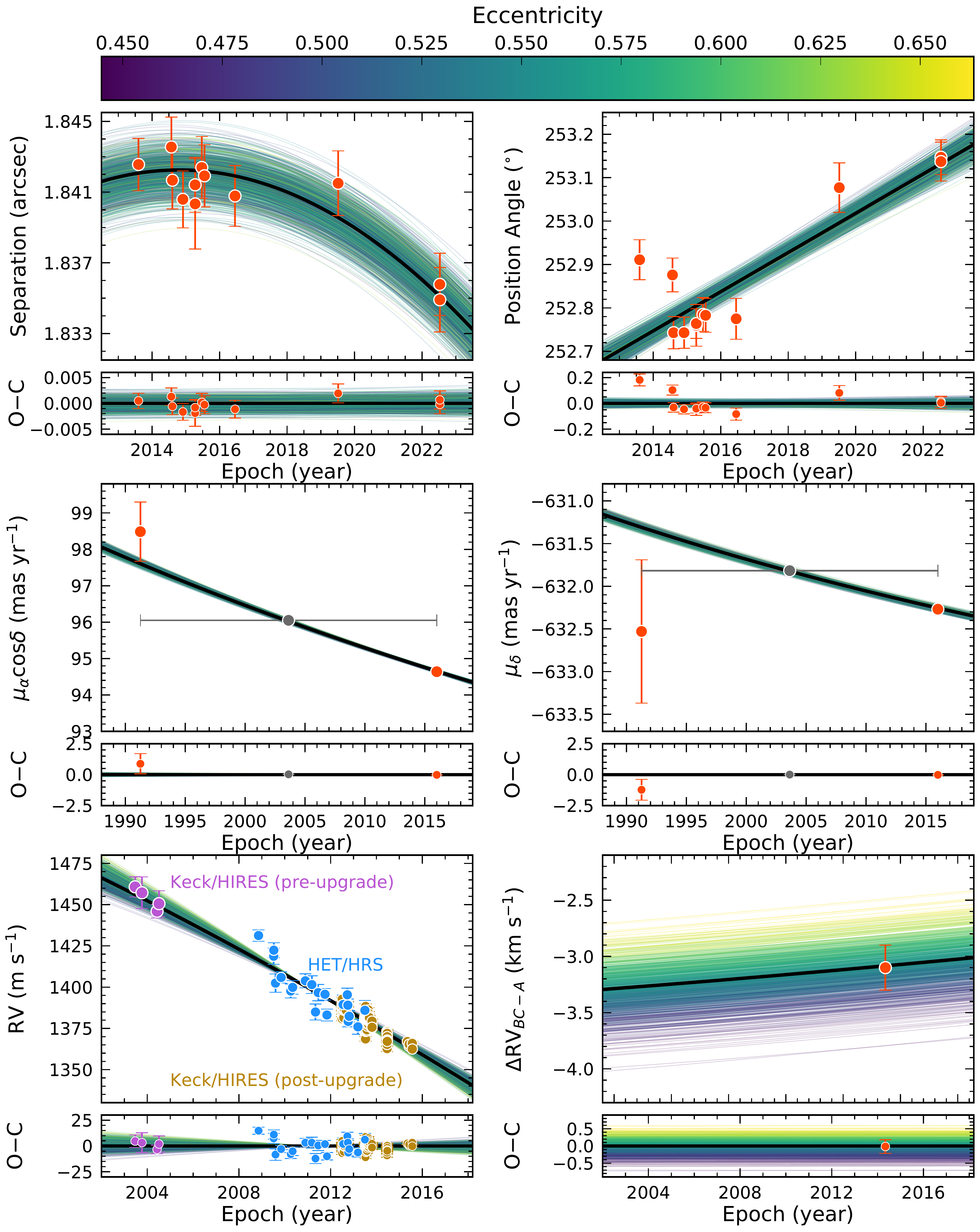}
\caption{Model predictions overlaid on the observed relative astrometry from Keck/NIRC2 (top), absolute astrometry from Hipparcos (J1991.25) and Gaia~EDR3 (J2016; middle), Kepler-444 A's multi-epoch RVs from HET/HRS and Keck/HIRES (bottom left), and the single-epoch BC-A relative RV from Keck/HIRES (bottom right). In each panel, we show the observed data (top) and residuals (bottom) using orange circles, except (1) the middle panels where we use grey circles to present the weighted-mean proper motion between Hipparcos and Gaia at J2003.625, the value that \texttt{orvara} uses to constrain the model-predicted proper motions of Kepler-444~A \citep{2021AJ....162..186B}, and (2) the bottom left panel where we use different colors to label RVs collected by different instruments. Predictions of 1000 randomly drawn orbits from the MCMC trials are overlaid in each panel color-coded by eccentricities. Predictions from the best-fit orbit are shown as black solid lines. }
\label{fig:orvara_combo}
\end{center}
\end{figure*}

Figure~\ref{fig:orvara_corner} presents the resulting parameter posteriors and the fitted sky-projected orbits of Kepler-444. We compare the observed relative astrometry, absolute astrometry, and radial velocities to model predictions in Figure~\ref{fig:orvara_combo}. The fitted and derived physical properties and uncertainties are listed in Table~\ref{tab:orbparams}. The entire set of MCMC chains of the orbit analysis presented here and those in the Appendix are accessible online.\footnote{\url{https://github.com/zjzhang42/Kepler_444_orbit_analysis} \label{footnote}}

{\tabletypesize{\scriptsize} 
\begin{deluxetable*}{llcccl}
\setlength{\tabcolsep}{10pt} 
\tablecaption{Orbit analysis of Kepler-444} \label{tab:orbparams} 
\tablehead{ \multicolumn{1}{l}{Parameter\tablenotemark{\scriptsize a}} &  \multicolumn{1}{l}{Unit} &  \multicolumn{1}{c}{Median$\pm1\sigma$} &  \multicolumn{1}{c}{$2\sigma$ Confidence Interval} &  \multicolumn{1}{c}{Best Fit} &  \multicolumn{1}{l}{Adopted Prior} } 
\startdata 
\multicolumn{6}{c}{Fitted Parameters} \\ 
\hline 
Mass of Kepler-444 A, $M_{A}$ &  $M_{\odot}$ & $0.75^{+0.03}_{-0.03}$  &  $(0.69, 0.81)$  &  $0.74$  &  $\mathcal{N}(\mu=0.75, \sigma^{2}=0.03^{2})$ \\  
Mass of Kepler-444 BC, $M_{BC}$ &  $M_{\odot}$ & $0.60^{+0.02}_{-0.02}$  &  $(0.57, 0.63)$  &  $0.60$  &  $1/M$ (log-flat) \\  
Semi-major axis, $a$ &  au & $52.2^{+3.3}_{-2.7}$  &  $(47.2, 59.4)$  &  $52.0$  &  $1/a$ (log-flat)  \\  
$\sqrt{e}\sin{\omega_{\star}}$ &  -- & $-0.55^{+0.03}_{-0.03}$  &  $(-0.61, -0.48)$  &  $-0.54$  &  Uniform  \\  
$\sqrt{e}\cos{\omega_{\star}}$ &  -- & $-0.50^{+0.08}_{-0.07}$  &  $(-0.63, -0.32)$  &  $-0.51$  &  Uniform  \\  
Inclination, $i$ &  degree & $85.4^{+0.3}_{-0.4}$  &  $(84.5, 86.0)$  &  $85.3$  &  $\sin{(i)}$ with $i \in [0, 180^{\circ}]$  \\  
PA of the ascending node, $\Omega$ &  degree & $250.7^{+0.2}_{-0.2}$  &  $(250.3, 251.1)$  &  $250.7$  &  Uniform  \\  
Mean longitude at J2010.0, $\lambda_{\rm ref,\star}$ &  degree & $338.9^{+1.7}_{-1.6}$  &  $(335.7, 342.3)$  &  $339.2$  &  Uniform  \\  
Parallax, $\varpi$ &  mas & $27.358^{+0.016}_{-0.016}$  &  $(27.325, 27.391)$  &  $27.361$  &  $\mathcal{N}(\mu=27.358, \sigma^{2}=0.013^{2})$  \\  
System Barycentric Proper Motion in RA, $\mu_{\alpha}\cos{(\delta)}$ &  mas~yr$^{-1}$ & $94.58^{+0.03}_{-0.03}$  &  $(94.52, 94.63)$  &  $94.59$  &  Uniform \\  
System Barycentric Proper Motion in DEC, $\mu_{\delta}$ &  mas~yr$^{-1}$ & $-631.61^{+0.04}_{-0.04}$  &  $(-631.68, -631.53)$  &  $-631.60$  &  Uniform  \\  
RV Jitter for HET/HRS, $\sigma_{\rm jit,HRS}$ &  m s$^{-1}$ & $6.2^{+1.6}_{-1.3}$  &  $(3.6, 9.3)$  &  $5.4$  &  $1/\sigma_{\rm jit,HRS}$ (log-flat)  \\  
RV zero point for HET/HRS, ZP$_{\rm HRS}$ &  m s$^{-1}$ & $1408^{+96}_{-94}$  &  $(1221, 1601)$  &  $1397$  &  Uniform  \\  
RV Jitter for post-upgrade HIRES, $\sigma_{\rm jit,post-HIRES}$ &  m s$^{-1}$ & $2.9^{+0.2}_{-0.2}$  &  $(2.5, 3.3)$  &  $2.9$  &  $1/\sigma_{\rm jit,post-HIRES}$ (log-flat)  \\  
RV zero point for post-upgrade HIRES, ZP$_{\rm post-HIRES}$ &  m s$^{-1}$ & $1390^{+95}_{-94}$  &  $(1203, 1583)$  &  $1379$  &  Uniform  \\  
RV Jitter for pre-upgrade HIRES, $\sigma_{\rm jit,pre-HIRES}$ &  m s$^{-1}$ & $0.0^{+0.7}_{-0.0}$  &  $(0.0, 5.3)$  &  $0.0$  &  $1/\sigma_{\rm jit,pre-HIRES}$ (log-flat)  \\  
RV zero point for pre-upgrade HIRES, ZP$_{\rm pre-HIRES}$ &  m s$^{-1}$ & $1463^{+96}_{-94}$  &  $(1275, 1657)$  &  $1451$  &  Uniform  \\  
\hline 
\multicolumn{6}{c}{Derived Parameters} \\ 
\hline 
Mass of Kepler-444, B $M_{B}$  &  $M_{\odot}$  &  $0.307^{+0.009}_{-0.008}$  &  $(0.290, 0.324)$  &  $0.308$  &  --  \\  
Logarithmic Mass of Kepler-444 B, $\log{(M_{B}/M_{\odot})}$  &  --  &  $-0.514^{+0.012}_{-0.012}$  &  $(-0.538, -0.489)$  &  $-0.511$  &  --  \\  
Mass of Kepler-444 C, $M_{C}$  &  $M_{\odot}$  &  $0.296^{+0.008}_{-0.008}$  &  $(0.280, 0.314)$  &  $0.297$  &  --  \\  
Logarithmic Mass of Kepler-444 C, $\log{(M_{C}/M_{\odot})}$  &  --  &  $-0.528^{+0.012}_{-0.012}$  &  $(-0.553, -0.504)$  &  $-0.527$  &  --  \\  
BC-to-A mass ratio, $M_{BC}/M_{A}$  &  --  &  $0.81^{+0.04}_{-0.04}$  &  $(0.73, 0.89)$  &  $0.80$  &  --  \\  
Eccentricity, $e$  &  --  &  $0.55^{+0.05}_{-0.05}$  &  $(0.46, 0.65)$  &  $0.55$  &  --  \\  
Argument of periastron, $\omega_{\star}$  &  degree  &  $227.3^{+6.5}_{-5.2}$  &  $(217.7, 241.7)$  &  $226.6$  &  --  \\  
Period, $P$  &  year  &  $324^{+31}_{-25}$  &  $(277, 396)$  &  $323$  &  --  \\  
Time of periastron, $T_{0}$\tablenotemark{\scriptsize b}  &  JD  &  $2537060^{+10881}_{-8533}$  &  $(2521634, 2562059)$  &  $2536428$  &  --  \\  
On-sky semi-major axis, $a\times\varpi$  &  mas  &  $1429^{+89}_{-74}$  &  $(1291, 1625)$  &  $1422$  &  --  \\  
Minimum A$-$BC separation, $a(1-e)$  &  au  &  $23^{+4}_{-4}$  &  $(17, 32)$  &  $23$  &  --  \\  
\enddata 
\tablenotetext{a}{Orbital parameters all correspond to Kepler-444 BC except for $a$, $\omega_{\star}$, and $\lambda_{\rm ref,\star}$. The first parameter corresponds to the system's (instead of individual components') semi-major axis, and the latter two parameters correspond to those of Kepler-444 A's orbit.}  
\tablenotetext{b}{$T_{0}$ is computed as $t_{\rm ref} - P \times (\lambda_{\rm ref,\star} - \omega_{\star})/360^{\circ}$, where $t_{\rm ref} = 2455197.5$~JD (i.e., epoch J2010.0).} 
\end{deluxetable*} 
}

Our analysis provides the latest characterization of Kepler-444 system's architecture based on a uniform re-analysis of all published data and new observations. Compared to \cite{2016ApJ...817...80D}, our newly derived semi-major axis of the system is $5\sigma$ larger $a = 52.2^{+3.3}_{-2.7}$~au (compared to $36.7^{+0.7}_{-0.9}$~au) and the eccentricity is $5.7\sigma$ smaller $e = 0.55 \pm 0.05$ (compared to $0.86 \pm 0.02$). These updates lead to a wider relative separation between A and BC during the periastron and imply a much larger size and mass of the truncated protoplanetary disk of Kepler-444~A (Section~\ref{subsec:disk_a}). The new inclination is consistent with the previous analysis although our updated value is $8.5$ times more precise $i = 85\dotdeg4^{+0\dotdeg3}_{-0\dotdeg4}$ (compared to $90\dotdeg4^{+3\dotdeg4}_{-3\dotdeg6}$). Therefore, we draw the same conclusion as \cite{2016ApJ...817...80D} that there is a possible orbital alignment between the stellar binary and transiting planets (Section~\ref{subsec:mutual_inclination}). Also, $\omega_{\star}$ is $\approx 120^{\circ}$ lower and $\Omega$ is $\approx 180^{\circ}$ higher, suggesting a different three-dimensional orientation of BC's barycentric orbit. 

We further measure the individual dynamical masses of Kepler-444~B and C for the first time, given that their total mass is well-constrained by our orbit analysis and the C-to-B mass ratio has been measured from multi-epoch absolute RVs of these two components (Section~\ref{subsec:deltarv}). We find $M_{B} = 0.307^{+0.009}_{-0.008}$~M$_{\odot}$ and $M_{C} = 0.296 \pm 0.008$~M$_{\odot}$, with $2\sigma$ intervals and best-fit values, are listed in Table~\ref{tab:orbparams}.

In addition, the relative RV between the primary and secondary components is not a common observable in the orbit analysis of stellar binaries, especially when the binary has a tight angular separation. To test the importance of this observable, we re-perform the orbit analysis by excluding the BC$-$A relative RV (Appendix~\ref{app:orbit_nodeltarv}). Without $\Delta{{\rm RV}_{BC-A}}$, we find the resulting parameter posteriors would be composed of two families of orbital solutions with similar shapes (e.g., semi-major axis and eccentricity) and line-of-sight inclinations, but completely different three-dimensional orientations. This analysis thus reveals the power of an even single-epoch relative RV between the primary and the secondary in order to precisely and accurately constrain the architecture of stellar binaries \citep[e.g.,][]{2020ApJ...894..115P}, especially when the secondary is near the apoapsis on a long-period orbit like Kepler-444.

\section{Discussion}
\label{sec:discussion}

\subsection{The Truncated Protoplanetary Disk of Kepler-444~A}
\label{subsec:disk_a}
 
The protoplanetary disk of Kepler-444~A was likely truncated by BC during the early evolutionary stages of this system (see \citealt{2021arXiv211206394Z} for a similar example). Therefore, the periastron separation between A and BC provides a boundary condition on the size and mass of this truncated disk. Based on an inferred periastron separation of $5.0^{+0.9}_{-1.0}$~au, \cite{2016ApJ...817...80D} estimated that A's disk likely had a radius of $2$~au, with a dust mass of $4$~M$_{\oplus}$ if the disk gas surface density follows the minimum mass solar nebula (MMSN). Their results imply that the primary star's disk would be too heavily depleted of solids to support the formation of five rocky planets unless the dust-to-planet conversion is very efficient or the disk surface density is slightly higher than the MMSN.

Here we re-examine the properties of Kepler-444~A's truncated disk using our new orbital parameters, which imply a $4.6 \pm 1.2$ times wider periastron separation between A and BC of $23 \pm 4$~au (Table~\ref{tab:orbparams}). \cite{1994ApJ...421..651A} performed an analytical study of disk-binary interactions and estimated the size of truncated circumprimary, circumsecondary, and circumbinary disks using the disk radius at which the resonant torque (from interactions between the disk and eccentric binary orbits) and the viscous torque (within the disk) are balanced. They computed truncated disk radii as functions of the mass ratio between binary components, the secondary's orbital eccentricity, and the disk viscosity (described by the Reynolds number $\mathcal{R}$), assuming the stellar binary and the disk are perfectly aligned. \cite{2019A&A...628A..95M} further expanded the numerical simulation results of \cite{1994ApJ...421..651A} into analytical functions, with the truncated radius of the circumprimary disk expressed as:
\begin{equation} \label{eq:trunc_disk}
R_{\rm disk,pri} = a \times \frac{0.49 \times q^{2/3}}{0.6 \times q^{2/3} + \ln{(1 + q^{1/3})}} \times \left( b \times e^{c} + 0.88 \mu^{0.01} \right)
\end{equation}
where $a$ is the system's semi-major axis, $e$ is the eccentricity of the secondary's orbit, $q = M_{\rm pri} / M_{\rm sec}$ is the primary-to-secondary mass ratio, and $\mu = M_{\rm sec} / (M_{\rm pri} + M_{\rm sec})$ is the secondary-to-total mass ratio. $b$ and $c$ are parameters that depend on $\mu$ and $\mathcal{R}$ \citep[see Table~C.1 in][]{2019A&A...628A..95M}. The truncated radius of the circumsecondary disk is expressed by the same equation with $q$ switched to the secondary-to-primary mass ratio $M_{\rm sec} / M_{\rm pri}$. 

Given that Kepler-444 has $\mu = 0.45 \pm 0.01$ based on our orbit analysis, we compute Kepler-444~A's disk radius using several combinations of $b$ and $c$ corresponding to $\mu = 0.4$ or $0.5$, and $\mathcal{R} = 10^{4}$, $10^{5}$, or $10^{6}$. The resulting disk radii span $7-9$~au with a typical uncertainty of $\approx 1$~au. In addition, the barycentric orbit of Kepler-444~BC and those of Kepler-444~A's transiting planets have mutual inclinations of at least $1\dotdeg6-4\dotdeg6$ (Section~\ref{subsec:mutual_inclination}), and if this misalignment is primordial, then Kepler-444 slightly deviates from the co-planarity assumption of \cite{1994ApJ...421..651A} embedded in Equation~\ref{eq:trunc_disk}. As suggested by \cite{2015ApJ...800...96L}, circumprimary or circumsecondary disks that are misaligned with the stellar binary orbit by $\psi$ can have systematically larger radii compared to those of aligned disks, as the resonant torque on the disk decays as $\cos^{8}{(\psi/2)}$ \citep[also see][]{2015MNRAS.452.2396M}. Therefore, we adopt a conservative truncation radius of $8$~au, which is 4 times larger than \cite{2016ApJ...817...80D}. 

We follow the same method as \cite{2016ApJ...817...80D} to estimate the potential reservoir of dust mass that resided in Kepler-444~A's disk. Specifically, we integrate an MMSN gas surface density of $\Sigma{(r)} = 1700 \times (r/{\rm 1\ au})^{3/2}$~g~cm$^{-2}$ \citep[][]{1977Ap&SS..51..153W, 1981PThPS..70...35H} using our estimated truncation disk radius and a dust-to-gas mass ratio of 1:300 (to incorporate the primary star's low metallicity of [Fe/H]$=-0.52\pm0.12$~dex; \citealt{2018A&A...612A..46M}). This leads to $500$~M$_\oplus$ or $1.6$~M$_{\rm Jup}$, implying a much larger potential mass reservoir of dust as compared to the value of $4$~M$_{\oplus}$ derived in \cite{2016ApJ...817...80D} under the same assumption of an MMSN disk. With a truncated disk radius of 2~au, \cite{2016ApJ...817...80D} suggested that a $\approx 20\times$ denser MMSN would be sufficient to explain the planet formation and such a disk would have a mass of $80-240$~M$_{\oplus}$ depending on the dust-to-gas mass ratio. We find these values are closer to our new estimate of the disk dust mass.

In addition to a more massive truncated disk of Kepler-444~A, we also update the estimates of planet masses. \cite{2016ApJ...817...80D} derived a total mass of 1.5~M$_{\oplus}$ for A's five planets based on these objects' measured radii and the \cite{2011ApJS..197....8L} mass-radius relation of $(M/M_{\oplus}) \sim (R/R_{\oplus})^{2.06}$. After this study, \cite{2017ApJ...838L..11M} used transit timing variation to directly constrain the photodynamical masses of Kepler-444~d and e to be $0.036^{+0.065}_{-0.020}$~M$_{\oplus}$ and $0.034^{+0.059}_{-0.019}$~M$_{\oplus}$, respectively. These measurements suggest that the planet d and e likely have water-rich or pure-rock compositions. These directly measured masses are 7 times smaller than those estimated by \cite{2016ApJ...817...80D}. This discrepancy is likely because the \cite{2011ApJS..197....8L} mass-radius relation was determined with Earth and Saturn, which have much larger densities than Kepler-444~A's planets. Using a mass-radius relation of $(R/R_{\oplus}) \sim (M/M_{\oplus})^{0.28}$ by \cite{2017ApJ...834...17C} for ``Terran worlds'' (with radii of $0.1-1$~R$_{\oplus}$), we find the predicted masses of d (0.104~M$_{\oplus}$) and e (0.115~M$_{\oplus}$) at their radii are about 3 times higher than the measured masses. Regardless, assuming Kepler-444~bcf planets all follow the \cite{2017ApJ...834...17C} Terran-world mass-radius relation, we compute their masses to be $0.039$~M$_{\oplus}$, $0.082$~$M_{\oplus}$, and $0.343$~M$_{\oplus}$, respectively, leading to a total mass of $0.53$~M$_{\oplus}$ for Kepler-444 planets. This total mass drops to $0.22$~M$_{\oplus}$ if the masses of b, c, and f are also 3 times smaller than the scaling-relation predictions as seen in d and e. 

With our updated estimates about the disk and planets' masses, Kepler-444's total planet mass within a given disk radius is well below the encompassed total disk dust mass. These planets' masses are still slightly higher than the predicted isolation mass of solids \citep[i.e., the maximum available mass reservoir needed for planets to undergo runaway accretion;][]{1987Icar...69..249L} at their currently observed locations in an MMSN disk \citep[e.g., see Figure~6 of][]{2016ApJ...817...80D}. Thus, it is likely that the disk surface density of Kepler-444~A is only slightly ($\approx 4\times$) higher than the MMSN. In addition, given that the truncated disk of Kepler-444~A is three orders of magnitudes more massive than the currently observed planet masses, it is possible that the Kepler-444~A's planets --- tightly packed within 0.1~au --- built their masses by accreting pebbles delivered from larger disk radii \citep[e.g.,][]{2014ApJ...780...53C, 2014ApJ...797...95L}, as discussed in \cite{2016ApJ...817...80D}. Therefore, we conclude that the previously noticed tension between Kepler-444~A's disk mass and its planet masses is now resolved by the new orbit analysis of this system.

\subsection{Mutual Inclinations between the Barycentric Orbit of Kepler-444~BC and Orbits of the Kepler-444~A Planets}
\label{subsec:mutual_inclination}
Mutual inclinations between Kepler-444~BC's barycentric orbit and the orbits of Kepler-444~A's planets provide valuable insight into the impact of stellar binaries on the formation and evolution of planets \citep[e.g.,][]{2019ApJ...883...22C, 2022MNRAS.512..648D, 2022AJ....163..207C}. Deriving this mutual inclination $\psi$ requires knowledge of the inclination $i$ and the position angle of the ascending node for orbits of both the outer binary($B$) and the inner planet ($p$):
\begin{equation}
\psi = \cos^{-1}{\big[ \cos{(i_{p})} \cos{(i_{B})} + \sin{(i_{p})} \sin{(i_{B})} \cos{(\Omega_{p} - \Omega_{B})} \big]}
\end{equation}
Given that $\Omega_{p}$ is usually unknown for transiting planets, only the minimum value of $\psi$ can be constrained as $|i_{p} - i_{B}|$ \citep[e.g.,][]{2017AJ....154..165B}. Our derived inclination of Kepler-444~BC is $85\dotdeg4^{+0\dotdeg3}_{-0\dotdeg4}$ and the observed inclinations of A's five planets span $87^{\circ} - 90^{\circ}$ \citep[][]{2015ApJ...799..170C}. Therefore, the true mutual inclination can be as small as $\psi = 1\dotdeg6-4\dotdeg6$. This result is consistent with \cite{2016ApJ...817...80D}, who derived $i_{B} = 90\dotdeg4^{+3\dotdeg4}_{-3\dotdeg6}$, leading a minimum $\psi = 0\dotdeg4-3\dotdeg4$.\footnote{In addition to the planet-binary mutual inclination, the inclination of Kepler-444~A's spin axis ($i_{A}$) was measured by \cite{2016ApJ...819...85C} using asteroseismology. Their inferred probability distribution of $i_{A}$ peaks at $90^{\circ}$, with wide $1\sigma$ and $2\sigma$ confidence intervals of $31\dotdeg3-90^{\circ}$ and $22\dotdeg7-90^{\circ}$, respectively. Given the large $i_{A}$ uncertainty and unknown sky-projected obliquities of the planets and BC, it remains unclear whether A's stellar spin axis, the planets' orbits, and BC's barycentric orbit are (mis-)aligned. Also, an alignment was suggested by \cite{1994AJ....107..306H} among a close ($\lesssim 30$~au) stellar binary's orbit and stellar spin axes of binary components, although recent studies found the existing data and precision are insufficient to assess the spin-orbit alignment of binaries \citep[e.g.,][]{2020A&A...642A.212J}. }

The mutual inclination could be significantly large if the orbital ascending node of BC's barycenter and those of planets have different position angles. However, if the orbital plane of BC-A and that of A's planets have large mutual inclinations, then the torque of the misaligned barycentric orbit of BC on the planets could cause the planets to precess as a rigid disk, which in turn would lead to cases where between none to all five planets are transiting along the line of sight \citep[e.g.,][]{2016ApJ...817...80D}. Therefore, it is likely that the orbital plane of BC-A and that of the planets are nearly aligned. As extensively discussed in \cite{2016ApJ...817...80D}, the potential coplanarity of the stellar and the planet orbits in the Kepler-444 system might be explained if they all formed within a large circumstellar disk, which fragmented to form the BC binary pair during the early evolutionary stages of the system and then subsequently formed the planets through core accretion at some later stage \citep[e.g.,][]{1989ApJ...347..959A, 1994MNRAS.271..999B, 2016ARA&A..54..271K, 2016Natur.538..483T, 2018AJ....155..160T, 2022arXiv220310066O}. Alternatively, the planet-binary coplanarity might also be a result of turbulent fragmentation, with BC having a primordial misalignment with A's protoplanetary disk. The disk could be torqued to precess by BC, with the energy dissipation driving the disk toward the aligned configuration  \citep[e.g.,][]{2000MNRAS.317..773B, 2012Natur.491..418B, 2018MNRAS.477.5207Z, 2022AJ....163..207C}.

It is noteworthy to compare our derived planet-binary mutual inclination of Kepler-444 with other observational evidence about about the statistical alignment between stellar binaries and their planets. \cite{2022AJ....163..207C} studied 67 host stars of candidate transiting planets identified by the Transiting Exoplanet Survey Satellite \citep[TESS;][]{2015JATIS...1a4003R}, which have outer stellar companions. They found that the measured orbital inclinations of the planet-host stellar binaries (particularly those with semi-major axes below 700~au) are preferentially closer to $i_{p}$ (assumed to be $90^{\circ}$), while the inclinations of binaries without planets follow an isotropic distribution. The overabundance of small $|i_{p} - i_{b}|$ (or $|90^{\circ} - i_{b}|$) in their samples thus points to a possible binary-planet alignment, given that these systems' $\Omega_{p}$ should be independent from $i_{p}$ or $i_{b}$. 

Also, \cite{2022MNRAS.512..648D} studied 45 planet-host stellar binaries and defined a metric $\gamma$, which is the angle between the secondary's on-sky orbital speed along the position angle (i.e., tangential) direction and that along the separation direction. Based on their definition, $\gamma$ is close to $0^{\circ}$ when the orbital motion along the tangential direction is zero, implying an edge-on orbit of the secondary and thereby a small $\psi$ between the binaries' and planets' orbits. The observed $\gamma$ distribution in their work is skewed toward $0^{\circ}$ and is best explained if orbits of stellar binaries and their planets are aligned within $30^{\circ}$ and if these binaries have uniformly distributed eccentricities within 0.1$-$0.8 \citep[similar to those of field binaries;][]{2010ApJS..190....1R}. 

In addition, \cite{2022AJ....163..160B} studied 168 host stars of TESS candidate transiting planets with outer stellar companions. Similar to \cite{2022MNRAS.512..648D}, they independently defined a metric $\gamma$ that measures the angle between a stellar binary's relative position vector and relative proper motion vector, as a probe of the planet-binary mutual inclination. Among a subset that host sub-Neptune or super-Earth planets (with planets' a$<1$~au and radii $\leqslant 4$~R$_{\oplus}$), they found $73^{+14}_{-20}\%$ of this set has planet-binary mutual inclinations of $35^{\circ} \pm 24^{\circ}$. However, among a subset that host close-in gas-giant planets (with planets' orbital periods $<10$~days and radii $> 4$~R$_{\oplus}$), which are not characteristic of the planets in Kepler-444, they found $65^{+20}_{-35}\%$ of these systems favor a perpendicular planet-binary mutual inclination of $89^{\circ} \pm 21^{\circ}$. 

Therefore, the potential alignment between BC's barycentric orbit and the orbits of A's planets in the Kepler-444 system generally lines up with those of statical samples. Direct constraints about the planet-binary mutual inclination have been rare in S-type planetary systems largely due to the unknown $\Omega_{p}$ of inner planets. In contrast, such measurements have been carried out for protoplanetary disks surrounding short-period ($P \leqslant 35$~days) spectroscopic binaries \citep[leading to small disk-binary mutual inclinations of $\leqslant 6^{\circ}$; e.g.,][]{2019ApJ...883...22C, 2021ApJ...912....6C}, as well as hierarchical stellar multiple systems \citep[e.g.,][]{2016MNRAS.455.4136B, 2016Natur.538..483T, 2018AJ....155..160T}.

\section{Conclusion}
\label{sec:conclusion}

We present the latest characterization of the architecture for the ancient ($\sim 11$~Gyr) Kepler-444 system, which is composed of a metal-poor ([Fe/H]$= -0.52 \pm 0.12$~dex) K0 primary star, Kepler-444~A, hosting 5 sub-Earth sized transiting planets, and a tight M-type spectroscopic binary, Kepler-444~BC. Combining our new observations and previously published data, we measure the system's relative astrometry, the primary star's muti-epoch RVs, and the BC$-$A relative RVs. We have also implemented the absolute astrometry and significant astrometric acceleration from Hipparcos and Gaia. 

Our work has provided significant updates to the orbital parameters of Kepler-444~BC's barycentric orbit compared to the previous work \citep[][]{2016ApJ...817...80D}, mainly because of our re-analysis of the BC$-$A relative RV and that our new observations have greatly extended the time baseline of the existing monitoring of the system's astrometry from 3 to 9 years. These updates include a $5\sigma$ larger semi-major axis ($a = 52.2^{+3.3}_{-2.7}$~au), a $5.7\sigma$ smaller eccentricity ($e = 0.55 \pm 0.05$), a more precise orbital inclination ($i = 85\dotdeg4^{+0\dotdeg3}_{-0\dotdeg4}$), a $\approx 120^{\circ}$ different argument of the primary star's periastron ($\omega_{\star} = 227\dotdeg3^{+6\dotdeg5}_{-5\dotdeg2}$), and a $\approx 180^{\circ}$ different position angle of the A-BC ascending node ($\Omega = 250\dotdeg7 \pm 0\dotdeg2$). We have also measured the first individual dynamical masses for the B ($0.307^{+0.009 }_{-0.008}$~M$_{\odot}$) and C ($0.296 \pm 0.008$~M$_{\odot}$) components. 

The updated $a$ and $e$ of Kepler-444~BC's barycentric orbit leads to a $4.6 \pm 1.2$~times wider relative separation between A and BC during periastron passage, suggesting the protoplanetary disk of Kepler-444~A was likely truncated to a radius of $\approx 8$~au by tidal interactions of BC, with a total dust mass of 500~M$_{\oplus}$ assuming an MMSN disk. We also update the total mass of Kepler-444~A's planets to be $0.53$~M$_{\oplus}$ by using the \cite{2017ApJ...834...17C} mass-radius relation and photodynamical mass measurements of Kepler-444~d and e \citep[][]{2017ApJ...838L..11M}. With our updated mass estimates of the truncated disk and planets, Kepler-444~A's five planets might have effectively built their masses via the accretion of pebbles delivered from larger disk radii if they formed in situ within a solid-depleted MMSN disk. This formation scenario was previously suggested by \cite{2016ApJ...817...80D}, under an assumption of very efficient dust-to-planet conversion or a much higher disk surface density than MMSN, given the tension between their lower mass estimates of the disk ($4$~M$_{\oplus}$) and higher mass estimates of the planets ($1.5$~M$_{\oplus}$). This tension is now resolved by the new orbit analysis.

The updated inclination of Kepler-444~BC's barycentric orbit leads to the same conclusion as \cite{2016ApJ...817...80D} that the orbital plane of A-BC and those of the planets are consistent with being aligned, with the planet-binary mutual inclination as small as $1\dotdeg6-4\dotdeg6$. A misalignment is possible if the ascending nodes of these planets' orbits do not line up with that of BC, but can cause situations where none to all five planets are transiting along the line of sight over the evolutionary history of this system. The coplanarity between the planets and the A-BC orbit might be explained if they all formed within a large circumstellar disk as extensively discussed in \cite{2016ApJ...817...80D} and lines up with recent statistical studies of planet-host stellar binaries. 

If we do not include the BC$-$A relative RV into our orbit analysis, then the resulting posteriors of orbital parameters are composed of two families of solutions, with comparable posterior probabilities, similar shapes, but completely different three-dimensional orientations. Therefore, for systems like Kepler-444, it is important to observe even single-epoch relative RVs between the primary and the secondary in order to precisely and accurately constrain the binary orbital architecture, especially when the secondary is near apoapsis on a long-period orbit.


\begin{acknowledgments}
Z. Z. thanks Benjamin Tofflemire and Songhu Wang for helpful discussions. Support for this work was provided by NASA through the NASA Hubble Fellowship grant HST-HF2-51522.001-A awarded by the Space Telescope Science Institute, which is operated by the Association of Universities for Research in Astronomy, Inc., for NASA, under contract NAS5-26555. B.P.B. acknowledges support from the National Science Foundation grant AST-1909209, NASA Exoplanet Research Program grant 20-XRP20$\_$2-0119, and the Alfred P. Sloan Foundation. This work was supported by a NASA Keck PI Data Award, administered by the NASA Exoplanet Science Institute. The data presented herein were in part obtained at the W. M. Keck Observatory, which is operated as a scientific partnership among the California Institute of Technology, the University of California and the National Aeronautics and Space Administration. The Observatory was made possible by the generous financial support of the W. M. Keck Foundation. The authors wish to recognize and acknowledge the very significant cultural role and reverence that the summit of Maunakea has always had within the indigenous Hawaiian community.  We are most fortunate to have the opportunity to conduct observations from this mountain. This research has made use of the Keck Observatory Archive (KOA), which is operated by the W. M. Keck Observatory and the NASA Exoplanet Science Institute (NExScI), under contract with the National Aeronautics and Space Administration. The Hobby-Eberly Telescope (HET) is a joint project of the University of Texas at Austin, the Pennsylvania State University, Ludwig-Maximilians-Universit\"{a}t M\"{u}nchen, and Georg-August-Universit\"{a}t G\"{o}ttingen. The HET is named in honor of its principal benefactors, William P. Hobby and Robert E. Eberly. This research has made use of the Spanish Virtual Observatory (\url{https://svo.cab.inta-csic.es}) project funded by MCIN/AEI/10.13039/501100011033/ through grant PID2020-112949GB-I00. For the purpose of open access, the author has applied a Creative Commons Attribution (CC BY) licence to any Author Accepted Manuscript version arising from this submission.
\end{acknowledgments}

\vspace{5mm}
\facilities{HET (HRS), Keck~II (NIRC2), Keck~I (HIRES)}
\software{\texttt{orvara} \citep[][]{2021AJ....162..186B}, \texttt{corner.py} \citep[][]{corner}, Astropy \citep{2013A&A...558A..33A, 2018AJ....156..123A}, IPython \citep{PER-GRA:2007}, Numpy \citep{numpy},  Scipy \citep{scipy}, Matplotlib \citep{Hunter:2007}.}

{\tabletypesize{\scriptsize} 
\begin{deluxetable*}{llccc}
\setlength{\tabcolsep}{10pt} 
\tablecaption{Orbit analysis of Kepler-444 with a broader $M_{A}$ prior of $0.75 \pm 0.15$~M$_{\odot}$} \label{tab:orbparams_uncertainprimmass} 
\tablehead{ \multicolumn{1}{l}{Parameter\tablenotemark{\scriptsize a}} &  \multicolumn{1}{l}{Unit} &  \multicolumn{1}{c}{Median$\pm1\sigma$} &  \multicolumn{1}{c}{$2\sigma$ Confidence Interval} &  \multicolumn{1}{c}{Best Fit} } 
\startdata 
\multicolumn{5}{c}{Fitted Parameters} \\ 
\hline 
Mass of Kepler-444 A, $M_{A}$ &  $M_{\odot}$ & $0.70^{+0.14}_{-0.14}$  &  $(0.42, 0.98)$  &  $0.76$  \\  
Mass of Kepler-444 BC, $M_{BC}$ &  $M_{\odot}$ & $0.60^{+0.02}_{-0.02}$  &  $(0.57, 0.63)$  &  $0.60$  \\  
Semi-major axis, $a$ &  au & $53.1^{+4.7}_{-3.5}$  &  $(47.0, 64.4)$  &  $52.3$  \\  
$\sqrt{e}\sin{\omega_{\star}}$ &  -- & $-0.55^{+0.04}_{-0.04}$  &  $(-0.62, -0.48)$  &  $-0.54$  \\  
$\sqrt{e}\cos{\omega_{\star}}$ &  -- & $-0.48^{+0.12}_{-0.09}$  &  $(-0.64, -0.20)$  &  $-0.50$  \\  
Inclination, $i$ &  degree & $85.4^{+0.3}_{-0.4}$  &  $(84.5, 86.0)$  &  $85.5$  \\  
PA of the ascending node, $\Omega$ &  degree & $250.7^{+0.2}_{-0.2}$  &  $(250.3, 251.1)$  &  $250.8$  \\  
Mean longitude at J2010.0, $\lambda_{\rm ref,\star}$ &  degree & $338.7^{+1.8}_{-1.8}$  &  $(335.2, 342.3)$  &  $338.8$  \\  
Parallax, $\varpi$ &  mas & $27.358^{+0.016}_{-0.016}$  &  $(27.325, 27.391)$  &  $27.363$  \\  
System Barycentric Proper Motion in RA, $\mu_{\alpha}\cos{(\delta)}$ &  mas~yr$^{-1}$ & $94.58^{+0.03}_{-0.03}$  &  $(94.52, 94.63)$  &  $94.57$ \\  
System Barycentric Proper Motion in DEC, $\mu_{\delta}$ &  mas~yr$^{-1}$ & $-631.58^{+0.09}_{-0.08}$  &  $(-631.72, -631.38)$  &  $-631.62$  \\  
RV Jitter for HET/HRS, $\sigma_{\rm jit,HRS}$ &  m s$^{-1}$ & $6.17^{+1.57}_{-1.34}$  &  $(3.63, 9.30)$  &  $6.24$  \\  
RV zero point for HET/HRS, ZP$_{\rm HRS}$ &  m s$^{-1}$ & $1458^{+198}_{-166}$  &  $(1151, 1897)$  &  $1408$   \\  
RV Jitter for post-upgrade HIRES, $\sigma_{\rm jit,post-HIRES}$ &  m s$^{-1}$ & $2.88^{+0.20}_{-0.19}$  &  $(2.52, 3.31)$  &  $2.83$  \\  
RV zero point for post-upgrade HIRES, ZP$_{\rm post-HIRES}$ &  m s$^{-1}$ & $1440^{+198}_{-165}$  &  $(1133, 1879)$  &  $1390$   \\  
RV Jitter for pre-upgrade HIRES, $\sigma_{\rm jit,pre-HIRES}$ &  m s$^{-1}$ & $0.01^{+0.72}_{-0.01}$  &  $(0.00, 5.33)$  &  $2.08$  \\  
RV zero point for pre-upgrade HIRES, ZP$_{\rm pre-HIRES}$ &  m s$^{-1}$ & $1513^{+198}_{-166}$  &  $(1205, 1952)$  &  $1462$   \\  
\hline 
\multicolumn{5}{c}{Derived Parameters} \\ 
\hline 
Mass of Kepler-444 B, $M_{B}$  &  $M_{\odot}$  &  $0.307^{+0.009}_{-0.008}$  &  $(0.290, 0.324)$  &  $0.303$  \\  
Logarithmic Mass of Kepler-444 B, $\log{(M_{B}/M_{\odot})}$  &  --  &  $-0.513^{+0.012}_{-0.012}$  &  $(-0.538, -0.489)$  &  $-0.519$  \\  
Mass of Kepler-444 C, $M_{C}$  &  $M_{\odot}$  &  $0.296^{+0.008}_{-0.008}$  &  $(0.280, 0.314)$  &  $0.290$  \\  
Logarithmic Mass of Kepler-444 C, $\log{(M_{C}/M_{\odot})}$  &  --  &  $-0.528^{+0.012}_{-0.012}$  &  $(-0.553, -0.504)$  &  $-0.538$  \\  
BC-to-A mass ratio, $M_{BC}/M_{A}$  &  --  &  $0.86^{+0.22}_{-0.15}$  &  $(0.61, 1.45)$  &  $0.79$  \\  
Eccentricity, $e$  &  --  &  $0.54^{+0.06}_{-0.06}$  &  $(0.42, 0.66)$  &  $0.55$  \\  
Argument of periastron, $\omega_{\star}$  &  degree  &  $228.8^{+9.4}_{-6.5}$  &  $(217.4, 252.2)$  &  $227.2$  \\  
Period, $P$  &  year  &  $338^{+62}_{-43}$  &  $(262, 496)$  &  $323$  \\  
Time of periastron, $T_{0}$\tablenotemark{\scriptsize b}  &  JD  &  $2541098^{+20089}_{-13351}$  &  $(2518285, 2594167)$  &  $2536702$  \\  
On-sky semi-major axis, $a\times\varpi$  &  mas  &  $1451^{+127}_{-94}$  &  $(1284, 1762)$  &  $1429$  \\  
Minimum A$-$BC separation, $a(1-e)$  &  au  &  $24.6^{+5.7}_{-4.6}$  &  $(16.2, 37.3)$  &  $23.7$  \\  
\enddata 
\tablenotetext{a}{Orbital parameters all correspond to Kepler-444 BC except for $a$, $\omega_{\star}$, and $\lambda_{\rm ref,\star}$. The first parameter corresponds to the system's (instead of individual components') semi-major axis, and the latter two parameters correspond to those of Kepler-444 A's orbit.}  
\tablenotetext{b}{$T_{0}$ is computed as $t_{\rm ref} - P \times (\lambda_{\rm ref,\star} - \omega_{\star})/360^{\circ}$, where $t_{\rm ref} = 2455197.5$~JD (i.e., epoch J2010.0).} 
\end{deluxetable*} 
}

\appendix

\begin{figure*}[t]
\begin{center}
\includegraphics[height=6.in]{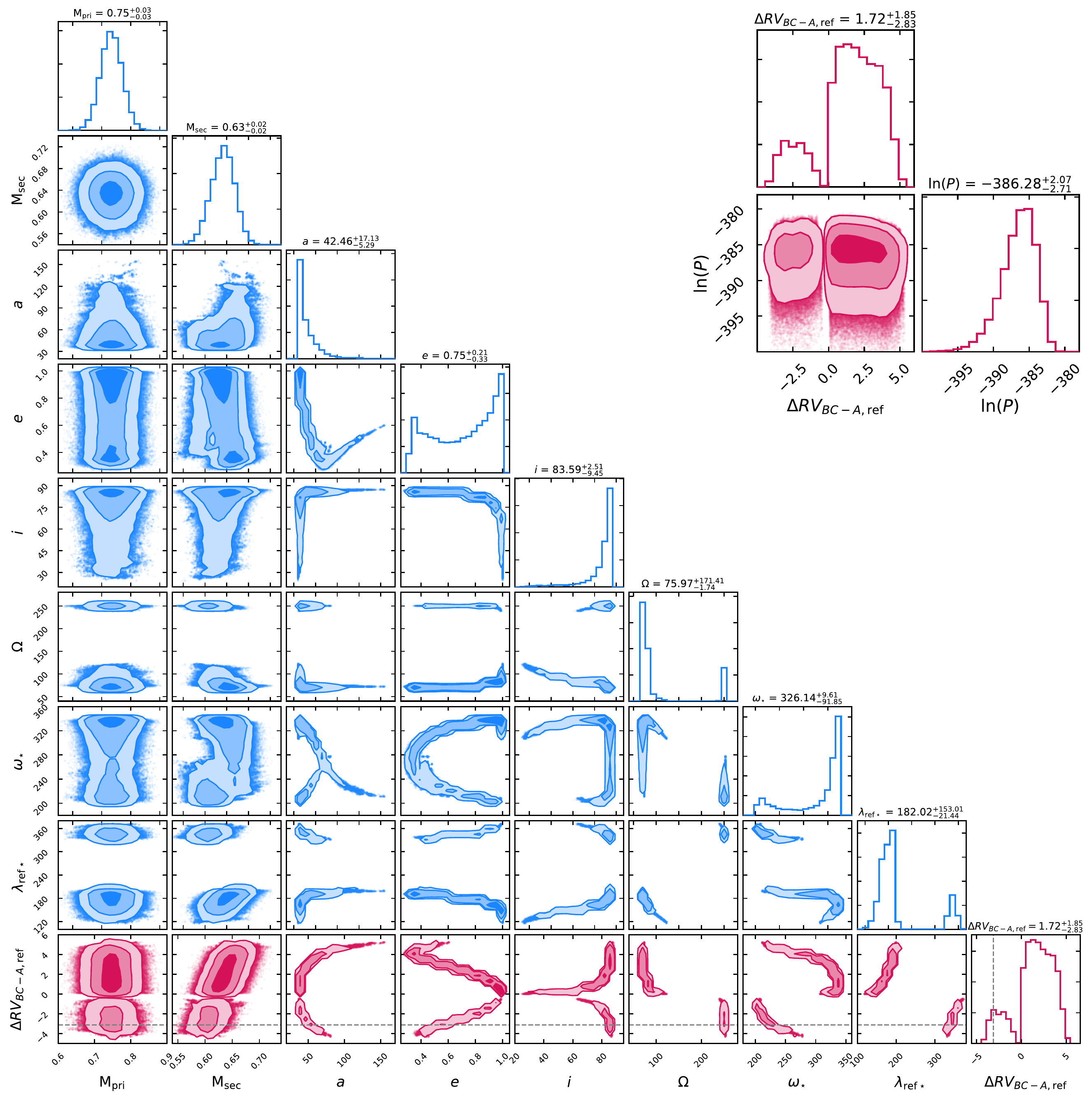}
\caption{Posteriors of our orbit analysis of Kepler-444 without including the observed BC$-$A relative radial velocity. The corresponding credible intervals and the best-fit values of these parameters are listed in Table~\ref{tab:orbparams_nodeltarv}. Fitted parameters are shown as the y-axis in the first eight rows of the corner plot and the y-axis of the last row presents $\Delta{{\rm RV}_{BC-A,\rm ref}}$, which is the calculated BC$-$A relative RV at the epoch of 2456783.1~JD (i.e., the mean epoch of our observed $\Delta{{\rm RV}_{BC-A}}$ value; Section~\ref{subsec:deltarv}). There are two families of orbital solutions, with one predicting positive $\Delta{{\rm RV}_{BC-A,\rm ref}}$ values and the other predicting negative $\Delta{{\rm RV}_{BC-A,\rm ref}}$ values. These two families of solutions have symmetric $a$, $e$, $i$ posteriors against $\Delta{{\rm RV}_{BC-A,\rm ref}} = 0$, but their $\Omega$, $\omega_{\star}$, and $\lambda_{\rm ref,\star}$ posteriors are bimodal, suggesting completely different three-dimensional orientations. The MCMC chains with $\Delta{{\rm RV}_{BC-A,\rm ref}} \approx -3.1$~km~s$^{-1}$ (horizontal dashed line) correspond to our fitted orbits in Section~\ref{sec:orbit}. At the top right, we show that the two families of solutions with different signs of $\Delta{{\rm RV}_{BC-A,\rm ref}}$ have comparable posterior probabilities, although their MCMC sample sizes are very different. }
\label{fig:orvara_corner_nodeltarv}
\end{center}
\end{figure*}

\begin{figure*}[t]
\begin{center}
\includegraphics[height=2.3in]{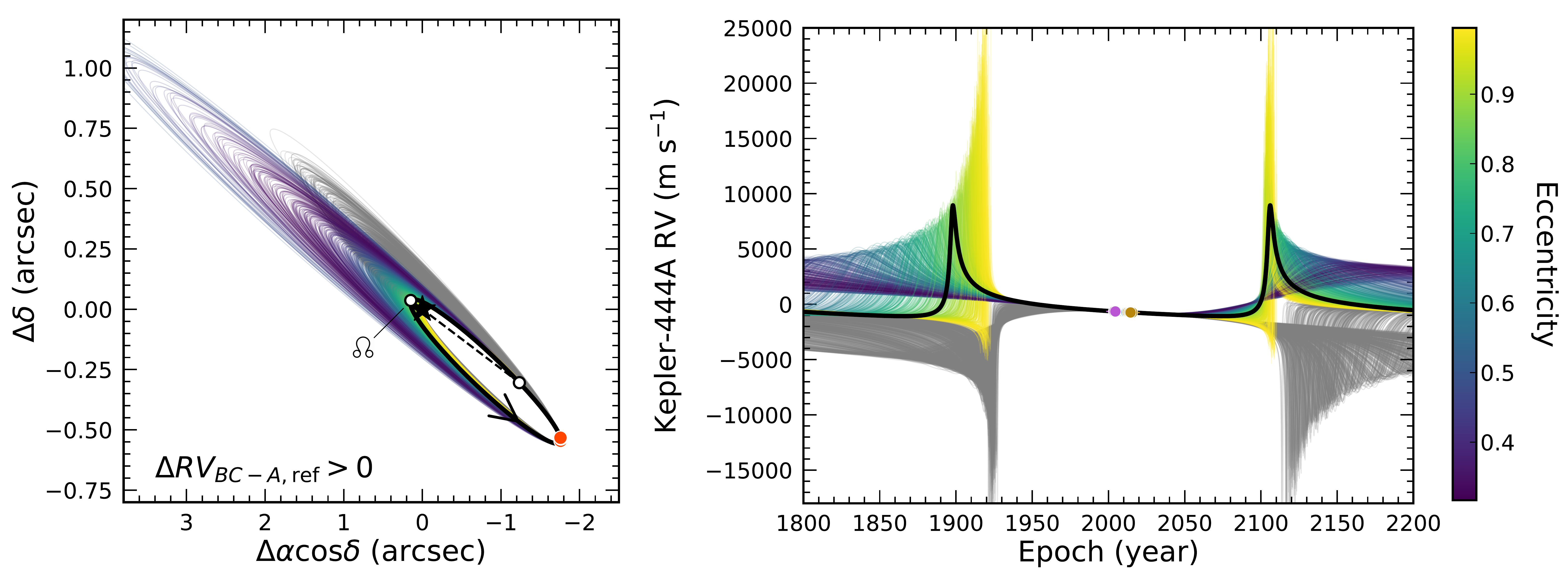}
\includegraphics[height=2.3in]{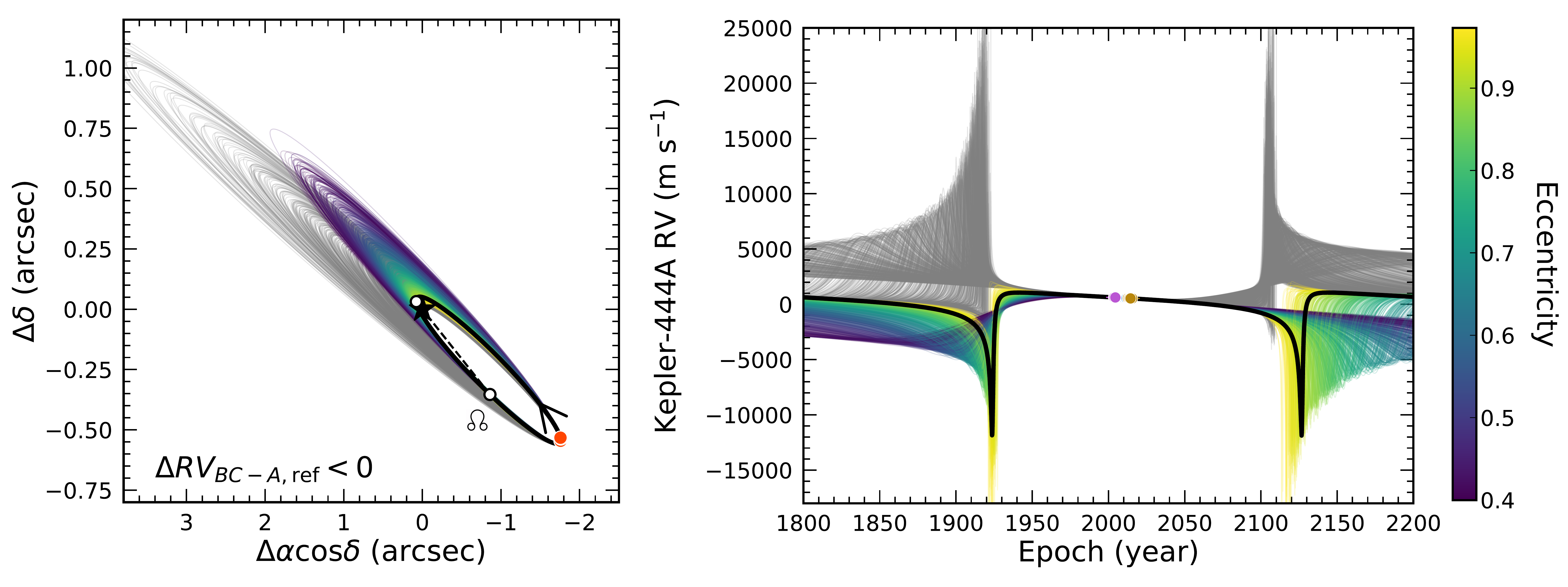}
\caption{{\it Top left}: Predicted relative astrometry between A and BC components of 1000 randomly drawn orbits from the MCMC chains with positive $\Delta{{\rm RV}_{BC-A,\rm ref}}$, color-coded by eccentricity. We overlay 1000 randomly drawn orbits from the MCMC chains with negative $\Delta{{\rm RV}_{BC-A,\rm ref}}$ and show them in grey. Similar to Figure~\ref{fig:orvara_corner}, we use the black solid line to show the best-fit orbit solution with $\Delta{{\rm RV}_{BC-A,\rm ref}} > 0$ and use two white circles to mark the ascending (labeled) and descending nodes, connected via a dashed line. We place Kepler-444~A (black star) at zero points and overlay the observed relative astrometry of BC (orange circles) that occupy the orbital arc at the bottom right. {\it Top right}: Predicted RVs of Kepler-444~A of the randomly drawn orbits (as shown in the top left panel). Orbits with positive $\Delta{{\rm RV}_{BC-A,\rm ref}}$ are color-coded by eccentricities while those with negative $\Delta{{\rm RV}_{BC-A,\rm ref}}$ are shown in grey, scaled to the same RV zero point. The black solid line shows predictions from the best-fit orbit, overlaid with the observed relative RVs of Kepler-444~A (purple, blue, and brown circles) color-coded by instruments in the same fashion as the bottom left panel in Figure~\ref{fig:orvara_combo}. {\it Bottom}: The same format as top, but we show the orbital solution with negative $\Delta{{\rm RV}_{BC-A,\rm ref}}$ in colors coded by eccentricities and those with positive $\Delta{{\rm RV}_{BC-A,\rm ref}}$ in grey. }
\label{fig:ra_dec_rv_positive_negative}
\end{center}
\end{figure*}

\section{Orbit Analysis of Kepler-444 with a broader $M_{A}$ Prior}
\label{app:orbit_broadprior}
Here we investigate the impact of our adopted prior of Kepler-444~A's mass on the derived dynamical mass and barycentric orbit of Kepler-444~BC. In Section~\ref{sec:orbit}, we set a Gaussian prior of $0.75 \pm 0.03$~M$_{\odot}$ for $M_{A}$ based on the recent estimate by \cite{2019A&A...630A.126B}, consistent with \cite{2015ApJ...799..170C}, \cite{2018A&A...612A..46M}, and \cite{2019A&A...622A.130B}, who derived $M_{A} = 0.76 \pm 0.04$~M$_{\odot}$, $0.76 \pm 0.01$~M$_{\odot}$, and $0.75 \pm 0.01$~M$_{\odot}$, respectively. The consistency of these measurements lines up with the expected small systematic error in mass ($\lesssim 5\%$) inferred from different evolution and pulsation codes \citep[e.g.,][]{2015MNRAS.452.2127S, 2021MNRAS.508.5864C, 2022ApJ...927...31T}. Nevertheless, to verify the robustness of our orbit analysis, here we assume a very conservative relative uncertainty of $20\%$ for the primary star's mass and adopt a broad $M_{A}$ prior of $0.75 \pm 0.15$~M$_{\odot}$ to perform the \texttt{orvara} analysis \citep[][]{2021AJ....162..186B} again with the same MCMC setup as in Section~\ref{sec:orbit}. Table~\ref{tab:orbparams_uncertainprimmass} presents our fitted and derived physical properties of Kepler-444. 

With a broader $M_{A}$ prior, we find the best-fit values and credible intervals of the following parameters remain nearly unchanged compared to our results in Section~\ref{sec:orbit}: individual masses of B and C components ($M_{B}$, $M_{C}$), eccentricity ($e$), inclination ($i$), position angle of the ascending node ($\Omega$), mean longitude of the primary star's orbit at epoch J2010.0 ($\lambda_{\rm ref,\star}$), the system's parallax ($\varpi$) and barycentric proper motion ($\mu_{\alpha}\cos{\delta}$ and $\mu_{\delta}$), and the RV jitter. The resulting RV ZPs are consistent within $0.3\sigma$ although those derived with a broader $M_{A}$ prior are systematically higher by $50$~m~s$^{-1}$. Also, the system's semi-major axis ($a$) and orbital period ($P$), the BC-to-A mass ratio, argument of the periastron of the primary star's orbit ($\omega_{\star}$), the time of periastron ($T_{0}$), as well as the relative separation between A and BC during periastron, all have consistent median values but $1.2-6\times$ larger uncertainties with a broader $M_{A}$ prior. These comparison results suggest that our orbital solution presented in Section~\ref{sec:orbit} is robust even with a very broad and conservative $M_{A}$ prior.

\section{Orbit Analysis of Kepler-444 without including the observed BC$-$A Relative RV}
\label{app:orbit_nodeltarv}
Here we use \texttt{orvara} to constrain the barycentric orbit and dynamical mass of Kepler-444~BC by using the system's relative astrometry, absolute astrometry, and the primary star's multi-epoch RVs, but excluding the observed BC$-$A relative RV. We set the same priors for free parameters and carry out the MCMC orbit analysis with the same number of temperatures, walkers, and steps as in Section~\ref{sec:orbit}. Along with the orbit fitting, we also calculate the BC$-$A relative RV at 2456783.1~JD (i.e., the mean epoch of our observed $\Delta{{\rm RV}_{BC-A}}$; Section~\ref{subsec:deltarv}) using the fitted parameters from each chain based on the following expression:
\begin{equation}
\begin{aligned}
\Delta{{\rm RV}_{BC-A}} &= - \frac{2\pi\sin{i} \sqrt{M_{A} + M_{BC}}}{\sqrt{a(1-e^{2})}} \left[\cos{(\nu + \omega_{\star})} + e \cos{(\omega_{\star})} \right]  \\
&\approx -4.74 \times \frac{2\pi\sin{i}}{\sqrt{1-e^{2}}} \left(\frac{M_{A} + M_{BC}}{M_{\odot}}\right)^{1/2} \left(\frac{a}{\rm au}\right)^{-1/2} \left[\cos{(\nu + \omega_{\star})} + e \cos{(\omega_{\star})} \right]\ {\rm km\ s^{-1}}
\end{aligned}
\end{equation}
where $\nu$ is the true anomaly. In following discussion, we use $\Delta{RV_{BC-A,\rm ref}}$ to note this calculated single-epoch BC$-$A relative RV.

As shown in Figure~\ref{fig:orvara_corner_nodeltarv}, the resulting posteriors from this reanalysis are composed to two families of solutions, with one predicting positive $\Delta{RV_{BC-A,\rm ref}}$ values (i.e., the BC component is moving away from us relative to A) and the other predicting negative $\Delta{RV_{BC-A,\rm ref}}$ values (i.e., the BC component is moving toward us relative to A). Both families of solutions produce the same posteriors in $M_{A}$ which is primarily constrained by the prior, but the ones with negative $\Delta{RV_{BC-A,\rm ref}}$ predict slightly lower masses for Kepler-444~BC. The posteriors of $a$, $e$, and $i$ are nearly symmetric against $\Delta{RV_{BC-A,\rm ref}} = 0$, with the eccentricity pushed toward an unphysical value of $\approx 1$ when $\Delta{RV_{BC-A,\rm ref}}$ is close to 0. Also, the distributions of $\omega_{\star}$, $\Omega$, and $\lambda_{\rm ref,\star}$ are bimodal, with distinct peak-to-peak separations of $\approx 120^{\circ}$, $180^{\circ}$, and $180^{\circ}$, respectively, suggesting the orbits' three-dimensional orientations of these two families of solutions are completely different (Figure~\ref{fig:ra_dec_rv_positive_negative}). In Table~\ref{tab:orbparams_nodeltarv}, we list the fitted and derived parameters and their uncertainties for each orbital solution. 

We note that only $17\%$ of the resulting MCMC chains produce negative $\Delta{RV_{BC-A,\rm ref}}$ values, and such unequal sample sizes between two families of orbital solutions are caused because the initial MCMC parameter values are closer to those producing a positive $\Delta{RV_{BC-A,\rm ref}}$. According to Figure~\ref{fig:orvara_corner_nodeltarv}, the computed posterior probabilities for each set of solutions are comparable. Therefore, our analysis reveals that with only the observed relative astrometry, absolute astrometry, and the primary star's RVs, we cannot distinguish between the two families of orbital solutions for the Kepler-444 system. Collecting Kepler-444~A's radial velocities while BC is near periapsis can help relieve the degeneracy, but this opportunity will not be available for another century (Figure~\ref{fig:ra_dec_rv_positive_negative}). In contrast, even a single epoch of the observed BC$-$A relative RV can efficiently break this degeneracy to precisely and accurately constrain the orbital parameters \citep[e.g.,][]{2020ApJ...894..115P}. Therefore, we encourage studies about the architectures of stellar binaries to consider observing the relative RV between the primary and the secondary stars, especially for systems similar to Kepler-444, where the secondary is near apoapsis on a long-period orbit.

{\tabletypesize{\scriptsize} 
\begin{deluxetable*}{llcccccccc}
\setlength{\tabcolsep}{5pt} 
\tablecaption{Orbit analysis of Kepler-444 without including the observed relative RV between BC and A Components} \label{tab:orbparams_nodeltarv} 
\tablehead{ \multicolumn{1}{l}{Parameter\tablenotemark{\scriptsize a}} &  \multicolumn{1}{l}{Unit} &  \multicolumn{1}{c}{} &  \multicolumn{3}{c}{Orbital Solution with positive $\Delta{RV_{BC-A, \rm ref}}$\tablenotemark{\scriptsize c}} &  \multicolumn{1}{c}{} &  \multicolumn{3}{c}{Orbital Solution with negative $\Delta{RV_{BC-A, \rm ref}}$\tablenotemark{\scriptsize c}} \\ 
\cline{4-6} \cline{8-10} \multicolumn{1}{l}{} &  \multicolumn{1}{l}{} &  \multicolumn{1}{c}{} &  \multicolumn{1}{c}{Median$\pm1\sigma$} &  \multicolumn{1}{c}{$2\sigma$ Confidence Interval} &  \multicolumn{1}{c}{Best Fit} &  \multicolumn{1}{c}{} &  \multicolumn{1}{c}{Median$\pm1\sigma$} &  \multicolumn{1}{c}{$2\sigma$ Confidence Interval} &  \multicolumn{1}{c}{Best Fit} } 
\startdata 
\multicolumn{10}{c}{Fitted Parameters} \\ 
\hline 
Mass of Kepler-444 A, $M_{A}$ & $M_{\odot}$ & & $0.75^{+0.03}_{-0.03}$  &  $(0.69, 0.81)$  &  $0.75$ & & $0.75^{+0.03}_{-0.03}$  &  $(0.69, 0.81)$  &  $0.75$  \\  
Mass of Kepler-444 BC, $M_{BC}$ & $M_{\odot}$ & & $0.64^{+0.02}_{-0.02}$  &  $(0.60, 0.67)$  &  $0.63$ & & $0.61^{+0.02}_{-0.02}$  &  $(0.57, 0.64)$  &  $0.61$  \\  
Semi-major axis, $a$ & au & & $41.8^{+18.5}_{-4.9}$  &  $(36.2, 94.5)$  &  $39.1$ & & $45.1^{+11.9}_{-6.0}$  &  $(37.3, 70.0)$  &  $38.2$  \\  
$\sqrt{e}\sin{\omega_{\star}}$ &  -- & & $-0.43^{+0.04}_{-0.11}$  &  $(-0.60, -0.36)$  &  $-0.38$ & & $-0.48^{+0.06}_{-0.11}$  &  $(-0.65, -0.38)$  &  $-0.41$  \\  
$\sqrt{e}\cos{\omega_{\star}}$ &  -- & & $0.76^{+0.12}_{-0.48}$  &  $(-0.37, 0.91)$  &  $0.85$ & & $-0.69^{+0.31}_{-0.16}$  &  $(-0.89, -0.08)$  &  $-0.87$  \\  
Inclination, $i$ &  degree & & $83.4^{+2.8}_{-11.4}$  &  $(40.3, 86.9)$  &  $81.2$ & & $84.2^{+1.6}_{-4.0}$  &  $(73.4, 86.4)$  &  $78.0$  \\  
PA of the ascending node, $\Omega$ &  degree & & $75.3^{+4.9}_{-1.1}$  &  $(73.8, 101.0)$  &  $76.1$ & & $250.2^{+0.7}_{-1.6}$  &  $(245.6, 251.2)$  &  $247.7$  \\  
Mean longitude at J2010.0, $\lambda_{\rm ref,\star}$ &  degree & & $177.0^{+15.4}_{-18.5}$  &  $(138.1, 198.1)$  &  $170.3$ & & $342.9^{+9.7}_{-7.0}$  &  $(2.1, 358.6)$  &  $358.0$  \\  
Parallax, $\varpi$ &  mas & & $27.358^{+0.016}_{-0.016}$  &  $(27.325, 27.391)$  &  $27.358$ & & $27.358^{+0.016}_{-0.016}$  &  $(27.325, 27.391)$  &  $27.353$  \\  
System Barycentric Proper Motion in RA, $\mu_{\alpha}\cos{(\delta)}$ &  mas~yr$^{-1}$ & & $94.57^{+0.03}_{-0.03}$  &  $(94.52, 94.63)$  &  $94.57$ & & $94.58^{+0.03}_{-0.03}$  &  $(94.52, 94.63)$  &  $94.57$ \\  
System Barycentric Proper Motion in DEC, $\mu_{\delta}$ &  mas~yr$^{-1}$ & & $-631.59^{+0.04}_{-0.04}$  &  $(-631.66, -631.51)$  &  $-631.60$ & & $-631.61^{+0.04}_{-0.04}$  &  $(-631.68, -631.53)$  &  $-631.59$ \\  
RV Jitter for HET/HRS, $\sigma_{\rm jit,HRS}$ &  m s$^{-1}$ & & $5.92^{+1.58}_{-1.35}$  &  $(3.35, 9.15)$  &  $5.81$ & & $6.20^{+1.57}_{-1.43}$  &  $(3.55, 9.35)$  &  $6.09$  \\  
RV zero point for HET/HRS, ZP$_{\rm HRS}$ &  m s$^{-1}$ & & $-967^{+647}_{-746}$  &  $(-2154, -55)$  &  $-708$  & & $1128^{+406}_{-448}$  &  $(406, 1760)$  &  $563$  \\  
RV Jitter for post-upgrade HIRES, $\sigma_{\rm jit,post-HIRES}$ &  m s$^{-1}$ & & $2.88^{+0.21}_{-0.19}$  &  $(2.52, 3.31)$  &  $2.88$ & & $2.89^{+0.19}_{-0.20}$  &  $(2.51, 3.28)$  &  $2.78$  \\  
RV zero point for post-upgrade HIRES, ZP$_{\rm post-HIRES}$ &  m s$^{-1}$ & & $-985^{+647}_{-746}$  &  $(-2173, -73)$  &  $-726$  & & $1110^{+406}_{-447}$  &  $(388, 1741)$  &  $545$  \\  
RV Jitter for pre-upgrade HIRES, $\sigma_{\rm jit,pre-HIRES}$ &  m s$^{-1}$ & & $0.01^{+0.66}_{-0.01}$  &  $(0.00, 5.03)$  &  $0.00$ & & $0.01^{+0.62}_{-0.01}$  &  $(0.00, 5.02)$  &  $0.12$  \\  
RV zero point for pre-upgrade HIRES, ZP$_{\rm pre-HIRES}$ &  m s$^{-1}$ & & $-904^{+645}_{-743}$  &  $(-2088, 4)$  &  $-647$   & & $-703^{+1284}_{-868}$  &  $(-2060, 1623)$  &  $-974$  \\  
\hline 
\multicolumn{10}{c}{Derived Parameters} \\ 
\hline 
BC-to-A mass ratio, $M_{BC}/M_{A}$  &  --  &  & $0.85^{+0.04}_{-0.04}$  &  $(0.77, 0.94)$  &  $0.84$  &  & $0.81^{+0.04}_{-0.04}$  &  $(0.74, 0.90)$  &  $0.82$  \\  
Eccentricity, $e$  &  --  &  & $0.76^{+0.20}_{-0.36}$  &  $(0.33, 0.99)$  &  $0.86$  &  & $0.70^{+0.19}_{-0.22}$  &  $(0.41, 0.96)$  &  $0.92$  \\  
Argument of periastron, $\omega_{\star}$  &  degree  & & $330.5^{+5.6}_{-33.2}$  &  $(234.2, 337.9)$  &  $335.6$  & & $214.7^{+21.9}_{-8.4}$  &  $(203.4, 263.2)$  &  $205.3$  \\  
Period, $P$  &  year  &  & $230^{+167}_{-38}$  &  $(185, 779)$  &  $208$  &  & $259^{+110}_{-50}$  &  $(194, 502)$  &  $202$  \\  
Time of periastron, $T_{0}$\tablenotemark{\scriptsize b}  &  JD  &  & $2490955^{+5090}_{-1274}$  &  $(2483815, 2498002)$  &  $2490081$  &  & $2515913^{+36893}_{-15739}$  &  $(2495193, 2602420)$  &  $2497876$  \\  
On-sky semi-major axis, $a\times\varpi$  &  mas  &  & $1144^{+506}_{-133}$  &  $(990, 2584)$  &  $1068$  &  & $1233^{+325}_{-164}$  &  $(1019, 1914)$  &  $1045$  \\  
Minimum A$-$BC separation, $a(1-e)$  &  au  &  & $10.0^{+27.5}_{-8.7}$  &  $(0.2, 56.8)$  &  $5.3$  &  & $13.5^{+15.9}_{-9.1}$  &  $(1.5, 40.9)$  &  $2.9$  \\  
BC$-$A relative RV at 2456783.1 JD $\Delta{RV_{BC-A,\rm ref}}$\tablenotemark{\scriptsize c}  &  km s$^{-1}$  &  & $2.2^{+1.6}_{-1.4}$  &  $(0.2, 4.6)$  &  $1.6$  &  & $-2.5^{+1.0}_{-0.9}$  &  $(-3.9, -0.8)$  &  $-1.2$  \\  
\enddata 
\tablenotetext{a}{Orbital parameters all correspond to Kepler-444 BC except for $a$, $\omega_{\star}$, and $\lambda_{\rm ref,\star}$. The first parameter corresponds to the system's (instead of individual components') semi-major axis, and the latter two parameters correspond to those of Kepler-444 A's orbit.}  
\tablenotetext{b}{$T_{0}$ is computed as $t_{\rm ref} - P \times (\lambda_{\rm ref,\star} - \omega_{\star})/360^{\circ}$, where $t_{\rm ref} = 2455197.5$~JD (i.e., epoch J2010.0).} 
\tablenotetext{c}{The $\Delta{RV_{BC-A,\rm ref}}$ is computed at 2456783.1 JD, the mean epoch of our measured $\Delta{RV_{BC-A}}$ value (Section~\ref{subsec:deltarv}).} 
\end{deluxetable*} 
}

\end{CJK*}

\clearpage
\bibliographystyle{aasjournal}
\bibliography{ms}{}

\begin{thebibliography}{}
\expandafter\ifx\csname natexlab\endcsname\relax\def\natexlab#1{#1}\fi
\providecommand{\url}[1]{\href{#1}{#1}}
\providecommand{\dodoi}[1]{doi:~\href{http://doi.org/#1}{\nolinkurl{#1}}}
\providecommand{\doeprint}[1]{\href{http://ascl.net/#1}{\nolinkurl{http://ascl.net/#1}}}
\providecommand{\doarXiv}[1]{\href{https://arxiv.org/abs/#1}{\nolinkurl{https://arxiv.org/abs/#1}}}

\bibitem[{{Adams} {et~al.}(1989){Adams}, {Ruden}, \&
  {Shu}}]{1989ApJ...347..959A}
{Adams}, F.~C., {Ruden}, S.~P., \& {Shu}, F.~H. 1989, \apj, 347, 959,
  \dodoi{10.1086/168187}

\bibitem[{{Adibekyan} {et~al.}(2012){Adibekyan}, {Delgado Mena}, {Sousa},
  {Santos}, {Israelian}, {Gonz{\'a}lez Hern{\'a}ndez}, {Mayor}, \&
  {Hakobyan}}]{2012A&A...547A..36A}
{Adibekyan}, V.~Z., {Delgado Mena}, E., {Sousa}, S.~G., {et~al.} 2012, \aap,
  547, A36, \dodoi{10.1051/0004-6361/201220167}

\bibitem[{{Arifyanto} \& {Fuchs}(2006)}]{2006A&A...449..533A}
{Arifyanto}, M.~I., \& {Fuchs}, B. 2006, \aap, 449, 533,
  \dodoi{10.1051/0004-6361:20054355}

\bibitem[{{Artymowicz} \& {Lubow}(1994)}]{1994ApJ...421..651A}
{Artymowicz}, P., \& {Lubow}, S.~H. 1994, \apj, 421, 651,
  \dodoi{10.1086/173679}

\bibitem[{{Astropy Collaboration} {et~al.}(2013){Astropy Collaboration},
  {Robitaille}, {Tollerud}, {Greenfield}, {Droettboom}, {Bray}, {Aldcroft},
  {Davis}, {Ginsburg}, {Price-Whelan}, {Kerzendorf}, {Conley}, {Crighton},
  {Barbary}, {Muna}, {Ferguson}, {Grollier}, {Parikh}, {Nair}, {Unther},
  {Deil}, {Woillez}, {Conseil}, {Kramer}, {Turner}, {Singer}, {Fox}, {Weaver},
  {Zabalza}, {Edwards}, {Azalee Bostroem}, {Burke}, {Casey}, {Crawford},
  {Dencheva}, {Ely}, {Jenness}, {Labrie}, {Lim}, {Pierfederici}, {Pontzen},
  {Ptak}, {Refsdal}, {Servillat}, \& {Streicher}}]{2013A&A...558A..33A}
{Astropy Collaboration}, {Robitaille}, T.~P., {Tollerud}, E.~J., {et~al.} 2013,
  \aap, 558, A33, \dodoi{10.1051/0004-6361/201322068}

\bibitem[{{Astropy Collaboration} {et~al.}(2018){Astropy Collaboration},
  {Price-Whelan}, {Sip{\H o}cz}, {G{\"u}nther}, {Lim}, {Crawford}, {Conseil},
  {Shupe}, {Craig}, {Dencheva}, {Ginsburg}, {VanderPlas}, {Bradley},
  {P{\'e}rez-Su{\'a}rez}, {de Val-Borro}, {Aldcroft}, {Cruz}, {Robitaille},
  {Tollerud}, {Ardelean}, {Babej}, {Bach}, {Bachetti}, {Bakanov}, {Bamford},
  {Barentsen}, {Barmby}, {Baumbach}, {Berry}, {Biscani}, {Boquien}, {Bostroem},
  {Bouma}, {Brammer}, {Bray}, {Breytenbach}, {Buddelmeijer}, {Burke},
  {Calderone}, {Cano Rodr{\'{\i}}guez}, {Cara}, {Cardoso}, {Cheedella},
  {Copin}, {Corrales}, {Crichton}, {D'Avella}, {Deil}, {Depagne}, {Dietrich},
  {Donath}, {Droettboom}, {Earl}, {Erben}, {Fabbro}, {Ferreira}, {Finethy},
  {Fox}, {Garrison}, {Gibbons}, {Goldstein}, {Gommers}, {Greco}, {Greenfield},
  {Groener}, {Grollier}, {Hagen}, {Hirst}, {Homeier}, {Horton}, {Hosseinzadeh},
  {Hu}, {Hunkeler}, {Ivezi{\'c}}, {Jain}, {Jenness}, {Kanarek}, {Kendrew},
  {Kern}, {Kerzendorf}, {Khvalko}, {King}, {Kirkby}, {Kulkarni}, {Kumar},
  {Lee}, {Lenz}, {Littlefair}, {Ma}, {Macleod}, {Mastropietro}, {McCully},
  {Montagnac}, {Morris}, {Mueller}, {Mumford}, {Muna}, {Murphy}, {Nelson},
  {Nguyen}, {Ninan}, {N{\"o}the}, {Ogaz}, {Oh}, {Parejko}, {Parley}, {Pascual},
  {Patil}, {Patil}, {Plunkett}, {Prochaska}, {Rastogi}, {Reddy Janga},
  {Sabater}, {Sakurikar}, {Seifert}, {Sherbert}, {Sherwood-Taylor}, {Shih},
  {Sick}, {Silbiger}, {Singanamalla}, {Singer}, {Sladen}, {Sooley},
  {Sornarajah}, {Streicher}, {Teuben}, {Thomas}, {Tremblay}, {Turner},
  {Terr{\'o}n}, {van Kerkwijk}, {de la Vega}, {Watkins}, {Weaver}, {Whitmore},
  {Woillez}, {Zabalza}, \& {Astropy Contributors}}]{2018AJ....156..123A}
{Astropy Collaboration}, {Price-Whelan}, A.~M., {Sip{\H o}cz}, B.~M., {et~al.}
  2018, \aj, 156, 123, \dodoi{10.3847/1538-3881/aabc4f}

\bibitem[{{Bailer-Jones} {et~al.}(2021){Bailer-Jones}, {Rybizki}, {Fouesneau},
  {Demleitner}, \& {Andrae}}]{2021AJ....161..147B}
{Bailer-Jones}, C.~A.~L., {Rybizki}, J., {Fouesneau}, M., {Demleitner}, M., \&
  {Andrae}, R. 2021, \aj, 161, 147, \dodoi{10.3847/1538-3881/abd806}

\bibitem[{{Bate} {et~al.}(2000){Bate}, {Bonnell}, {Clarke}, {Lubow}, {Ogilvie},
  {Pringle}, \& {Tout}}]{2000MNRAS.317..773B}
{Bate}, M.~R., {Bonnell}, I.~A., {Clarke}, C.~J., {et~al.} 2000, \mnras, 317,
  773, \dodoi{10.1046/j.1365-8711.2000.03648.x}

\bibitem[{{Batygin}(2012)}]{2012Natur.491..418B}
{Batygin}, K. 2012, \nat, 491, 418, \dodoi{10.1038/nature11560}

\bibitem[{{Behmard} {et~al.}(2022){Behmard}, {Dai}, \&
  {Howard}}]{2022AJ....163..160B}
{Behmard}, A., {Dai}, F., \& {Howard}, A.~W. 2022, \aj, 163, 160,
  \dodoi{10.3847/1538-3881/ac53a7}

\bibitem[{{Bellinger} {et~al.}(2019){Bellinger}, {Hekker}, {Angelou},
  {Stokholm}, \& {Basu}}]{2019A&A...622A.130B}
{Bellinger}, E.~P., {Hekker}, S., {Angelou}, G.~C., {Stokholm}, A., \& {Basu},
  S. 2019, \aap, 622, A130, \dodoi{10.1051/0004-6361/201834461}

\bibitem[{{Bensby} {et~al.}(2014){Bensby}, {Feltzing}, \&
  {Oey}}]{2014A&A...562A..71B}
{Bensby}, T., {Feltzing}, S., \& {Oey}, M.~S. 2014, \aap, 562, A71,
  \dodoi{10.1051/0004-6361/201322631}

\bibitem[{{Bonavita} {et~al.}(2022){Bonavita}, {Fontanive}, {Gratton},
  {Mu{\v{z}}i{\'c}}, {Desidera}, {Mesa}, {Biller}, {Scholz}, {Sozzetti}, \&
  {Squicciarini}}]{2022MNRAS.513.5588B}
{Bonavita}, M., {Fontanive}, C., {Gratton}, R., {et~al.} 2022, \mnras, 513,
  5588, \dodoi{10.1093/mnras/stac1250}

\bibitem[{{Bonnell} \& {Bate}(1994)}]{1994MNRAS.271..999B}
{Bonnell}, I.~A., \& {Bate}, M.~R. 1994, \mnras, 271, 999,
  \dodoi{10.1093/mnras/271.4.999}

\bibitem[{{Borkovits} {et~al.}(2016){Borkovits}, {Hajdu}, {Sztakovics},
  {Rappaport}, {Levine}, {B{\'\i}r{\'o}}, \& {Klagyivik}}]{2016MNRAS.455.4136B}
{Borkovits}, T., {Hajdu}, T., {Sztakovics}, J., {et~al.} 2016, \mnras, 455,
  4136, \dodoi{10.1093/mnras/stv2530}

\bibitem[{{Bovy} {et~al.}(2009){Bovy}, {Hogg}, \&
  {Roweis}}]{2009ApJ...700.1794B}
{Bovy}, J., {Hogg}, D.~W., \& {Roweis}, S.~T. 2009, \apj, 700, 1794,
  \dodoi{10.1088/0004-637X/700/2/1794}

\bibitem[{{Bowler} {et~al.}(2017){Bowler}, {Kraus}, {Bryan}, {Knutson},
  {Brogi}, {Rizzuto}, {Mace}, {Vanderburg}, {Liu}, {Hillenbrand}, \&
  {Cieza}}]{2017AJ....154..165B}
{Bowler}, B.~P., {Kraus}, A.~L., {Bryan}, M.~L., {et~al.} 2017, \aj, 154, 165,
  \dodoi{10.3847/1538-3881/aa88bd}

\bibitem[{{Bowler} {et~al.}(2018){Bowler}, {Dupuy}, {Endl}, {Cochran},
  {MacQueen}, {Fulton}, {Petigura}, {Howard}, {Hirsch}, {Kratter}, {Crepp},
  {Biller}, {Johnson}, \& {Wittenmyer}}]{2018AJ....155..159B}
{Bowler}, B.~P., {Dupuy}, T.~J., {Endl}, M., {et~al.} 2018, \aj, 155, 159,
  \dodoi{10.3847/1538-3881/aab2a6}

\bibitem[{{Bowler} {et~al.}(2021{\natexlab{a}}){Bowler}, {Cochran}, {Endl},
  {Franson}, {Brandt}, {Dupuy}, {MacQueen}, {Kratter}, {Mawet}, \&
  {Ruane}}]{2021AJ....161..106B}
{Bowler}, B.~P., {Cochran}, W.~D., {Endl}, M., {et~al.} 2021{\natexlab{a}},
  \aj, 161, 106, \dodoi{10.3847/1538-3881/abd243}

\bibitem[{{Bowler} {et~al.}(2021{\natexlab{b}}){Bowler}, {Endl}, {Cochran},
  {MacQueen}, {Crepp}, {Doppmann}, {Dulz}, {Brandt}, {Mirek Brandt}, {Li},
  {Dupuy}, {Franson}, {Kratter}, {Morley}, \& {Zhou}}]{2021ApJ...913L..26B}
{Bowler}, B.~P., {Endl}, M., {Cochran}, W.~D., {et~al.} 2021{\natexlab{b}},
  \apjl, 913, L26, \dodoi{10.3847/2041-8213/abfec8}

\bibitem[{{Brandt}(2018)}]{2018ApJS..239...31B}
{Brandt}, T.~D. 2018, \apjs, 239, 31, \dodoi{10.3847/1538-4365/aaec06}

\bibitem[{{Brandt}(2021)}]{2021ApJS..254...42B}
---. 2021, \apjs, 254, 42, \dodoi{10.3847/1538-4365/abf93c}

\bibitem[{{Brandt} {et~al.}(2021){Brandt}, {Dupuy}, {Li}, {Brandt}, {Zeng},
  {Michalik}, {Bardalez Gagliuffi}, \& {Raposo-Pulido}}]{2021AJ....162..186B}
{Brandt}, T.~D., {Dupuy}, T.~J., {Li}, Y., {et~al.} 2021, \aj, 162, 186,
  \dodoi{10.3847/1538-3881/ac042e}

\bibitem[{{Brewer} {et~al.}(2016){Brewer}, {Fischer}, {Valenti}, \&
  {Piskunov}}]{2016ApJS..225...32B}
{Brewer}, J.~M., {Fischer}, D.~A., {Valenti}, J.~A., \& {Piskunov}, N. 2016,
  \apjs, 225, 32, \dodoi{10.3847/0067-0049/225/2/32}

\bibitem[{{Brewer} {et~al.}(2018){Brewer}, {Wang}, {Fischer}, \&
  {Foreman-Mackey}}]{2018ApJ...867L...3B}
{Brewer}, J.~M., {Wang}, S., {Fischer}, D.~A., \& {Foreman-Mackey}, D. 2018,
  \apjl, 867, L3, \dodoi{10.3847/2041-8213/aae710}

\bibitem[{{Buldgen} {et~al.}(2019){Buldgen}, {Farnir}, {Pezzotti},
  {Eggenberger}, {Salmon}, {Montalban}, {Ferguson}, {Khan}, {Bourrier},
  {Rendle}, {Meynet}, {Miglio}, \& {Noels}}]{2019A&A...630A.126B}
{Buldgen}, G., {Farnir}, M., {Pezzotti}, C., {et~al.} 2019, \aap, 630, A126,
  \dodoi{10.1051/0004-6361/201936126}

\bibitem[{{Butler} {et~al.}(2017){Butler}, {Vogt}, {Laughlin}, {Burt},
  {Rivera}, {Tuomi}, {Teske}, {Arriagada}, {Diaz}, {Holden}, \&
  {Keiser}}]{2017AJ....153..208B}
{Butler}, R.~P., {Vogt}, S.~S., {Laughlin}, G., {et~al.} 2017, \aj, 153, 208,
  \dodoi{10.3847/1538-3881/aa66ca}

\bibitem[{{Campante} {et~al.}(2015){Campante}, {Barclay}, {Swift}, {Huber},
  {Adibekyan}, {Cochran}, {Burke}, {Isaacson}, {Quintana}, {Davies}, {Silva
  Aguirre}, {Ragozzine}, {Riddle}, {Baranec}, {Basu}, {Chaplin},
  {Christensen-Dalsgaard}, {Metcalfe}, {Bedding}, {Handberg}, {Stello},
  {Brewer}, {Hekker}, {Karoff}, {Kolbl}, {Law}, {Lundkvist}, {Miglio}, {Rowe},
  {Santos}, {Van Laerhoven}, {Arentoft}, {Elsworth}, {Fischer}, {Kawaler},
  {Kjeldsen}, {Lund}, {Marcy}, {Sousa}, {Sozzetti}, \&
  {White}}]{2015ApJ...799..170C}
{Campante}, T.~L., {Barclay}, T., {Swift}, J.~J., {et~al.} 2015, \apj, 799,
  170, \dodoi{10.1088/0004-637X/799/2/170}

\bibitem[{{Campante} {et~al.}(2016){Campante}, {Lund}, {Kuszlewicz}, {Davies},
  {Chaplin}, {Albrecht}, {Winn}, {Bedding}, {Benomar}, {Bossini}, {Handberg},
  {Santos}, {Van Eylen}, {Basu}, {Christensen-Dalsgaard}, {Elsworth}, {Hekker},
  {Hirano}, {Huber}, {Karoff}, {Kjeldsen}, {Lundkvist}, {North}, {Silva
  Aguirre}, {Stello}, \& {White}}]{2016ApJ...819...85C}
{Campante}, T.~L., {Lund}, M.~N., {Kuszlewicz}, J.~S., {et~al.} 2016, \apj,
  819, 85, \dodoi{10.3847/0004-637X/819/1/85}

\bibitem[{{Chatterjee} \& {Tan}(2014)}]{2014ApJ...780...53C}
{Chatterjee}, S., \& {Tan}, J.~C. 2014, \apj, 780, 53,
  \dodoi{10.1088/0004-637X/780/1/53}

\bibitem[{{Chen} \& {Kipping}(2017)}]{2017ApJ...834...17C}
{Chen}, J., \& {Kipping}, D. 2017, \apj, 834, 17,
  \dodoi{10.3847/1538-4357/834/1/17}

\bibitem[{{Christian} {et~al.}(2022){Christian}, {Vanderburg}, {Becker},
  {Yahalomi}, {Pearce}, {Zhou}, {Collins}, {Kraus}, {Stassun}, {de Beurs},
  {Ricker}, {Vanderspek}, {Latham}, {Winn}, {Seager}, {Jenkins}, {Abe},
  {Agabi}, {Amado}, {Baker}, {Barkaoui}, {Benkhaldoun}, {Benni}, {Berberian},
  {Berlind}, {Bieryla}, {Esparza-Borges}, {Bowen}, {Brown}, {Buchhave},
  {Burke}, {Buttu}, {Cadieux}, {Caldwell}, {Charbonneau}, {Chazov},
  {Chimaladinne}, {Collins}, {Combs}, {Conti}, {Crouzet}, {de Leon},
  {Deljookorani}, {Diamond}, {Doyon}, {Dragomir}, {Dransfield}, {Essack},
  {Evans}, {Fukui}, {Gan}, {Esquerdo}, {Gillon}, {Girardin}, {Guerra},
  {Guillot}, {K. Habich}, {Henriksen}, {Hoch}, {Isogai}, {Jehin}, {Jensen},
  {Johnson}, {Livingston}, {Kielkopf}, {Kim}, {Kawauchi}, {Krushinsky},
  {Kunzle}, {Laloum}, {Leger}, {Lewin}, {Mallia}, {Massey}, {Mori}, {McLeod},
  {M{\'e}karnia}, {Mireles}, {Mishevskiy}, {Tamura}, {Murgas}, {Narita},
  {Naves}, {Nelson}, {Osborn}, {Palle}, {Parviainen}, {Plavchan}, {Pozuelos},
  {Rabus}, {Relles}, {Rodr{\'\i}guez L{\'o}pez}, {Quinn}, {Schmider},
  {Schlieder}, {Schwarz}, {Shporer}, {Sibbald}, {Srdoc}, {Stibbards},
  {Stickler}, {Suarez}, {Stockdale}, {Tan}, {Terada}, {Triaud}, {Tronsgaard},
  {Waalkes}, {Wang}, {Watanabe}, {Wenceslas}, {Wingham}, {Wittrock}, \&
  {Ziegler}}]{2022AJ....163..207C}
{Christian}, S., {Vanderburg}, A., {Becker}, J., {et~al.} 2022, \aj, 163, 207,
  \dodoi{10.3847/1538-3881/ac517f}

\bibitem[{{Cunha} {et~al.}(2021){Cunha}, {Roxburgh}, {Aguirre B{\o}rsen-Koch},
  {Ball}, {Basu}, {Chaplin}, {Goupil}, {Nsamba}, {Ong}, {Reese}, {Verma},
  {Belkacem}, {Campante}, {Christensen-Dalsgaard}, {Clara}, {Deheuvels},
  {Monteiro}, {Noll}, {Ouazzani}, {R{\o}rsted}, {Stokholm}, \&
  {Winther}}]{2021MNRAS.508.5864C}
{Cunha}, M.~S., {Roxburgh}, I.~W., {Aguirre B{\o}rsen-Koch}, V., {et~al.} 2021,
  \mnras, 508, 5864, \dodoi{10.1093/mnras/stab2886}

\bibitem[{{Currie} {et~al.}(2020){Currie}, {Brandt}, {Kuzuhara}, {Chilcote},
  {Guyon}, {Marois}, {Groff}, {Lozi}, {Vievard}, {Sahoo}, {Deo}, {Jovanovic},
  {Martinache}, {Wagner}, {Dupuy}, {Wahl}, {Letawsky}, {Li}, {Zeng}, {Brandt},
  {Michalik}, {Grady}, {Janson}, {Knapp}, {Kwon}, {Lawson}, {McElwain},
  {Uyama}, {Wisniewski}, \& {Tamura}}]{2020ApJ...904L..25C}
{Currie}, T., {Brandt}, T.~D., {Kuzuhara}, M., {et~al.} 2020, \apjl, 904, L25,
  \dodoi{10.3847/2041-8213/abc631}

\bibitem[{{Czekala} {et~al.}(2019){Czekala}, {Chiang}, {Andrews}, {Jensen},
  {Torres}, {Wilner}, {Stassun}, \& {Macintosh}}]{2019ApJ...883...22C}
{Czekala}, I., {Chiang}, E., {Andrews}, S.~M., {et~al.} 2019, \apj, 883, 22,
  \dodoi{10.3847/1538-4357/ab287b}

\bibitem[{{Czekala} {et~al.}(2021){Czekala}, {Ribas}, {Cuello}, {Chiang},
  {Mac{\'\i}as}, {Duch{\^e}ne}, {Andrews}, \&
  {Espaillat}}]{2021ApJ...912....6C}
{Czekala}, I., {Ribas}, {\'A}., {Cuello}, N., {et~al.} 2021, \apj, 912, 6,
  \dodoi{10.3847/1538-4357/abebe3}

\bibitem[{{Doyle} {et~al.}(2011){Doyle}, {Carter}, {Fabrycky}, {Slawson},
  {Howell}, {Winn}, {Orosz}, {P{\v{r}}sa}, {Welsh}, {Quinn}, {Latham},
  {Torres}, {Buchhave}, {Marcy}, {Fortney}, {Shporer}, {Ford}, {Lissauer},
  {Ragozzine}, {Rucker}, {Batalha}, {Jenkins}, {Borucki}, {Koch}, {Middour},
  {Hall}, {McCauliff}, {Fanelli}, {Quintana}, {Holman}, {Caldwell}, {Still},
  {Stefanik}, {Brown}, {Esquerdo}, {Tang}, {Furesz}, {Geary}, {Berlind},
  {Calkins}, {Short}, {Steffen}, {Sasselov}, {Dunham}, {Cochran}, {Boss},
  {Haas}, {Buzasi}, \& {Fischer}}]{2011Sci...333.1602D}
{Doyle}, L.~R., {Carter}, J.~A., {Fabrycky}, D.~C., {et~al.} 2011, Science,
  333, 1602, \dodoi{10.1126/science.1210923}

\bibitem[{{Duch{\^e}ne} \& {Kraus}(2013)}]{2013ARA&A..51..269D}
{Duch{\^e}ne}, G., \& {Kraus}, A. 2013, \araa, 51, 269,
  \dodoi{10.1146/annurev-astro-081710-102602}

\bibitem[{{Dupuy} {et~al.}(2016){Dupuy}, {Kratter}, {Kraus}, {Isaacson},
  {Mann}, {Ireland}, {Howard}, \& {Huber}}]{2016ApJ...817...80D}
{Dupuy}, T.~J., {Kratter}, K.~M., {Kraus}, A.~L., {et~al.} 2016, \apj, 817, 80,
  \dodoi{10.3847/0004-637X/817/1/80}

\bibitem[{{Dupuy} {et~al.}(2022){Dupuy}, {Kraus}, {Kratter}, {Rizzuto}, {Mann},
  {Huber}, \& {Ireland}}]{2022MNRAS.512..648D}
{Dupuy}, T.~J., {Kraus}, A.~L., {Kratter}, K.~M., {et~al.} 2022, \mnras, 512,
  648, \dodoi{10.1093/mnras/stac306}

\bibitem[{{Duquennoy} \& {Mayor}(1991)}]{1991A&A...248..485D}
{Duquennoy}, A., \& {Mayor}, M. 1991, \aap, 248, 485

\bibitem[{{Endl} {et~al.}(2000){Endl}, {K{\"u}rster}, \&
  {Els}}]{2000A&A...362..585E}
{Endl}, M., {K{\"u}rster}, M., \& {Els}, S. 2000, \aap, 362, 585

\bibitem[{{Fischer} \& {Marcy}(1992)}]{1992ApJ...396..178F}
{Fischer}, D.~A., \& {Marcy}, G.~W. 1992, \apj, 396, 178,
  \dodoi{10.1086/171708}

\bibitem[{{Fontanive} {et~al.}(2019){Fontanive}, {Mu{\v{z}}i{\'c}}, {},
  {Bonavita}, \& {Biller}}]{2019MNRAS.490.1120F}
{Fontanive}, C., {Mu{\v{z}}i{\'c}}, {}, K., {Bonavita}, M., \& {Biller}, B.
  2019, \mnras, 490, 1120, \dodoi{10.1093/mnras/stz2587}

\bibitem[{Foreman-Mackey(2016)}]{corner}
Foreman-Mackey, D. 2016, The Journal of Open Source Software, 1, 24,
  \dodoi{10.21105/joss.00024}

\bibitem[{{Foreman-Mackey} {et~al.}(2013){Foreman-Mackey}, {Hogg}, {Lang}, \&
  {Goodman}}]{2013PASP..125..306F}
{Foreman-Mackey}, D., {Hogg}, D.~W., {Lang}, D., \& {Goodman}, J. 2013, \pasp,
  125, 306, \dodoi{10.1086/670067}

\bibitem[{{Fouqu{\'e}} {et~al.}(2018){Fouqu{\'e}}, {Moutou}, {Malo},
  {Martioli}, {Lim}, {Rajpurohit}, {Artigau}, {Delfosse}, {Donati},
  {Forveille}, {Morin}, {Allard}, {Delage}, {Doyon}, {H{\'e}brard}, \&
  {Neves}}]{2018MNRAS.475.1960F}
{Fouqu{\'e}}, P., {Moutou}, C., {Malo}, L., {et~al.} 2018, \mnras, 475, 1960,
  \dodoi{10.1093/mnras/stx3246}

\bibitem[{{Franson} {et~al.}(2022){Franson}, {Bowler}, {Brandt}, {Dupuy},
  {Tran}, {Brandt}, {Li}, \& {Kraus}}]{2022AJ....163...50F}
{Franson}, K., {Bowler}, B.~P., {Brandt}, T.~D., {et~al.} 2022, \aj, 163, 50,
  \dodoi{10.3847/1538-3881/ac35e8}

\bibitem[{{Gaia Collaboration} {et~al.}(2016){Gaia Collaboration}, {Prusti},
  {de Bruijne}, {Brown}, {Vallenari}, {Babusiaux}, {Bailer-Jones}, {Bastian},
  {Biermann}, {Evans}, {Eyer}, {Jansen}, {Jordi}, {Klioner}, {Lammers},
  {Lindegren}, {Luri}, {Mignard}, {Milligan}, {Panem}, {Poinsignon},
  {Pourbaix}, {Randich}, {Sarri}, {Sartoretti}, {Siddiqui}, {Soubiran},
  {Valette}, {van Leeuwen}, {Walton}, {Aerts}, {Arenou}, {Cropper}, {Drimmel},
  {H{\o}g}, {Katz}, {Lattanzi}, {O'Mullane}, {Grebel}, {Holland}, {Huc},
  {Passot}, {Bramante}, {Cacciari}, {Casta{\~n}eda}, {Chaoul}, {Cheek}, {De
  Angeli}, {Fabricius}, {Guerra}, {Hern{\'a}ndez}, {Jean-Antoine-Piccolo},
  {Masana}, {Messineo}, {Mowlavi}, {Nienartowicz}, {Ord{\'o}{\~n}ez-Blanco},
  {Panuzzo}, {Portell}, {Richards}, {Riello}, {Seabroke}, {Tanga},
  {Th{\'e}venin}, {Torra}, {Els}, {Gracia-Abril}, {Comoretto},
  {Garcia-Reinaldos}, {Lock}, {Mercier}, {Altmann}, {Andrae}, {Astraatmadja},
  {Bellas-Velidis}, {Benson}, {Berthier}, {Blomme}, {Busso}, {Carry},
  {Cellino}, {Clementini}, {Cowell}, {Creevey}, {Cuypers}, {Davidson}, {De
  Ridder}, {de Torres}, {Delchambre}, {Dell'Oro}, {Ducourant}, {Fr{\'e}mat},
  {Garc{\'\i}a-Torres}, {Gosset}, {Halbwachs}, {Hambly}, {Harrison}, {Hauser},
  {Hestroffer}, {Hodgkin}, {Huckle}, {Hutton}, {Jasniewicz}, {Jordan},
  {Kontizas}, {Korn}, {Lanzafame}, {Manteiga}, {Moitinho}, {Muinonen},
  {Osinde}, {Pancino}, {Pauwels}, {Petit}, {Recio-Blanco}, {Robin}, {Sarro},
  {Siopis}, {Smith}, {Smith}, {Sozzetti}, {Thuillot}, {van Reeven}, {Viala},
  {Abbas}, {Abreu Aramburu}, {Accart}, {Aguado}, {Allan}, {Allasia},
  {Altavilla}, {{\'A}lvarez}, {Alves}, {Anderson}, {Andrei}, {Anglada Varela},
  {Antiche}, {Antoja}, {Ant{\'o}n}, {Arcay}, {Atzei}, {Ayache}, {Bach},
  {Baker}, {Balaguer-N{\'u}{\~n}ez}, {Barache}, {Barata}, {Barbier}, {Barblan},
  {Baroni}, {Barrado y Navascu{\'e}s}, {Barros}, {Barstow}, {Becciani},
  {Bellazzini}, {Bellei}, {Bello Garc{\'\i}a}, {Belokurov}, {Bendjoya},
  {Berihuete}, {Bianchi}, {Bienaym{\'e}}, {Billebaud}, {Blagorodnova},
  {Blanco-Cuaresma}, {Boch}, {Bombrun}, {Borrachero}, {Bouquillon}, {Bourda},
  {Bouy}, {Bragaglia}, {Breddels}, {Brouillet}, {Br{\"u}semeister},
  {Bucciarelli}, {Budnik}, {Burgess}, {Burgon}, {Burlacu}, {Busonero}, {Buzzi},
  {Caffau}, {Cambras}, {Campbell}, {Cancelliere}, {Cantat-Gaudin}, {Carlucci},
  {Carrasco}, {Castellani}, {Charlot}, {Charnas}, {Charvet}, {Chassat},
  {Chiavassa}, {Clotet}, {Cocozza}, {Collins}, {Collins}, {Costigan}, {Crifo},
  {Cross}, {Crosta}, {Crowley}, {Dafonte}, {Damerdji}, {Dapergolas}, {David},
  {David}, {De Cat}, {de Felice}, {de Laverny}, {De Luise}, {De March}, {de
  Martino}, {de Souza}, {Debosscher}, {del Pozo}, {Delbo}, {Delgado},
  {Delgado}, {di Marco}, {Di Matteo}, {Diakite}, {Distefano}, {Dolding}, {Dos
  Anjos}, {Drazinos}, {Dur{\'a}n}, {Dzigan}, {Ecale}, {Edvardsson}, {Enke},
  {Erdmann}, {Escolar}, {Espina}, {Evans}, {Eynard Bontemps}, {Fabre},
  {Fabrizio}, {Faigler}, {Falc{\~a}o}, {Farr{\`a}s Casas}, {Faye}, {Federici},
  {Fedorets}, {Fern{\'a}ndez-Hern{\'a}ndez}, {Fernique}, {Fienga}, {Figueras},
  {Filippi}, {Findeisen}, {Fonti}, {Fouesneau}, {Fraile}, {Fraser}, {Fuchs},
  {Furnell}, {Gai}, {Galleti}, {Galluccio}, {Garabato}, {Garc{\'\i}a-Sedano},
  {Gar{\'e}}, {Garofalo}, {Garralda}, {Gavras}, {Gerssen}, {Geyer}, {Gilmore},
  {Girona}, {Giuffrida}, {Gomes}, {Gonz{\'a}lez-Marcos},
  {Gonz{\'a}lez-N{\'u}{\~n}ez}, {Gonz{\'a}lez-Vidal}, {Granvik}, {Guerrier},
  {Guillout}, {Guiraud}, {G{\'u}rpide}, {Guti{\'e}rrez-S{\'a}nchez}, {Guy},
  {Haigron}, {Hatzidimitriou}, {Haywood}, {Heiter}, {Helmi}, {Hobbs},
  {Hofmann}, {Holl}, {Holland}, {Hunt}, {Hypki}, {Icardi}, {Irwin}, {Jevardat
  de Fombelle}, {Jofr{\'e}}, {Jonker}, {Jorissen}, {Julbe}, {Karampelas},
  {Kochoska}, {Kohley}, {Kolenberg}, {Kontizas}, {Koposov}, {Kordopatis},
  {Koubsky}, {Kowalczyk}, {Krone-Martins}, {Kudryashova}, {Kull}, {Bachchan},
  {Lacoste-Seris}, {Lanza}, {Lavigne}, {Le Poncin-Lafitte}, {Lebreton},
  {Lebzelter}, {Leccia}, {Leclerc}, {Lecoeur-Taibi}, {Lemaitre}, {Lenhardt},
  {Leroux}, {Liao}, {Licata}, {Lindstr{\o}m}, {Lister}, {Livanou}, {Lobel},
  {L{\"o}ffler}, {L{\'o}pez}, {Lopez-Lozano}, {Lorenz}, {Loureiro},
  {MacDonald}, {Magalh{\~a}es Fernandes}, {Managau}, {Mann}, {Mantelet},
  {Marchal}, {Marchant}, {Marconi}, {Marie}, {Marinoni}, {Marrese},
  {Marschalk{\'o}}, {Marshall}, {Mart{\'\i}n-Fleitas}, {Martino}, {Mary},
  {Matijevi{\v{c}}}, {Mazeh}, {McMillan}, {Messina}, {Mestre}, {Michalik},
  {Millar}, {Miranda}, {Molina}, {Molinaro}, {Molinaro}, {Moln{\'a}r},
  {Moniez}, {Montegriffo}, {Monteiro}, {Mor}, {Mora}, {Morbidelli}, {Morel},
  {Morgenthaler}, {Morley}, {Morris}, {Mulone}, {Muraveva}, {Musella},
  {Narbonne}, {Nelemans}, {Nicastro}, {Noval}, {Ord{\'e}novic},
  {Ordieres-Mer{\'e}}, {Osborne}, {Pagani}, {Pagano}, {Pailler}, {Palacin},
  {Palaversa}, {Parsons}, {Paulsen}, {Pecoraro}, {Pedrosa}, {Pentik{\"a}inen},
  {Pereira}, {Pichon}, {Piersimoni}, {Pineau}, {Plachy}, {Plum}, {Poujoulet},
  {Pr{\v{s}}a}, {Pulone}, {Ragaini}, {Rago}, {Rambaux}, {Ramos-Lerate},
  {Ranalli}, {Rauw}, {Read}, {Regibo}, {Renk}, {Reyl{\'e}}, {Ribeiro},
  {Rimoldini}, {Ripepi}, {Riva}, {Rixon}, {Roelens}, {Romero-G{\'o}mez},
  {Rowell}, {Royer}, {Rudolph}, {Ruiz-Dern}, {Sadowski}, {Sagrist{\`a}
  Sell{\'e}s}, {Sahlmann}, {Salgado}, {Salguero}, {Sarasso}, {Savietto},
  {Schnorhk}, {Schultheis}, {Sciacca}, {Segol}, {Segovia}, {Segransan},
  {Serpell}, {Shih}, {Smareglia}, {Smart}, {Smith}, {Solano}, {Solitro},
  {Sordo}, {Soria Nieto}, {Souchay}, {Spagna}, {Spoto}, {Stampa}, {Steele},
  {Steidelm{\"u}ller}, {Stephenson}, {Stoev}, {Suess}, {S{\"u}veges}, {Surdej},
  {Szabados}, {Szegedi-Elek}, {Tapiador}, {Taris}, {Tauran}, {Taylor},
  {Teixeira}, {Terrett}, {Tingley}, {Trager}, {Turon}, {Ulla}, {Utrilla},
  {Valentini}, {van Elteren}, {Van Hemelryck}, {van Leeuwen}, {Varadi},
  {Vecchiato}, {Veljanoski}, {Via}, {Vicente}, {Vogt}, {Voss}, {Votruba},
  {Voutsinas}, {Walmsley}, {Weiler}, {Weingrill}, {Werner}, {Wevers},
  {Whitehead}, {Wyrzykowski}, {Yoldas}, {{\v{Z}}erjal}, {Zucker}, {Zurbach},
  {Zwitter}, {Alecu}, {Allen}, {Allende Prieto}, {Amorim},
  {Anglada-Escud{\'e}}, {Arsenijevic}, {Azaz}, {Balm}, {Beck}, {Bernstein},
  {Bigot}, {Bijaoui}, {Blasco}, {Bonfigli}, {Bono}, {Boudreault}, {Bressan},
  {Brown}, {Brunet}, {Bunclark}, {Buonanno}, {Butkevich}, {Carret}, {Carrion},
  {Chemin}, {Ch{\'e}reau}, {Corcione}, {Darmigny}, {de Boer}, {de Teodoro}, {de
  Zeeuw}, {Delle Luche}, {Domingues}, {Dubath}, {Fodor}, {Fr{\'e}zouls},
  {Fries}, {Fustes}, {Fyfe}, {Gallardo}, {Gallegos}, {Gardiol}, {Gebran},
  {Gomboc}, {G{\'o}mez}, {Grux}, {Gueguen}, {Heyrovsky}, {Hoar}, {Iannicola},
  {Isasi Parache}, {Janotto}, {Joliet}, {Jonckheere}, {Keil}, {Kim},
  {Klagyivik}, {Klar}, {Knude}, {Kochukhov}, {Kolka}, {Kos}, {Kutka}, {Lainey},
  {LeBouquin}, {Liu}, {Loreggia}, {Makarov}, {Marseille}, {Martayan},
  {Martinez-Rubi}, {Massart}, {Meynadier}, {Mignot}, {Munari}, {Nguyen},
  {Nordlander}, {Ocvirk}, {O'Flaherty}, {Olias Sanz}, {Ortiz}, {Osorio},
  {Oszkiewicz}, {Ouzounis}, {Palmer}, {Park}, {Pasquato}, {Peltzer}, {Peralta},
  {P{\'e}turaud}, {Pieniluoma}, {Pigozzi}, {Poels}, {Prat}, {Prod'homme},
  {Raison}, {Rebordao}, {Risquez}, {Rocca-Volmerange}, {Rosen}, {Ruiz-Fuertes},
  {Russo}, {Sembay}, {Serraller Vizcaino}, {Short}, {Siebert}, {Silva},
  {Sinachopoulos}, {Slezak}, {Soffel}, {Sosnowska}, {Strai{\v{z}}ys}, {ter
  Linden}, {Terrell}, {Theil}, {Tiede}, {Troisi}, {Tsalmantza}, {Tur},
  {Vaccari}, {Vachier}, {Valles}, {Van Hamme}, {Veltz}, {Virtanen}, {Wallut},
  {Wichmann}, {Wilkinson}, {Ziaeepour}, \& {Zschocke}}]{2016A&A...595A...1G}
{Gaia Collaboration}, {Prusti}, T., {de Bruijne}, J.~H.~J., {et~al.} 2016,
  \aap, 595, A1, \dodoi{10.1051/0004-6361/201629272}

\bibitem[{{Gaia Collaboration} {et~al.}(2021){Gaia Collaboration}, {Brown},
  {Vallenari}, {Prusti}, {de Bruijne}, {Babusiaux}, {Biermann}, {Creevey},
  {Evans}, {Eyer}, {Hutton}, {Jansen}, {Jordi}, {Klioner}, {Lammers},
  {Lindegren}, {Luri}, {Mignard}, {Panem}, {Pourbaix}, {Randich}, {Sartoretti},
  {Soubiran}, {Walton}, {Arenou}, {Bailer-Jones}, {Bastian}, {Cropper},
  {Drimmel}, {Katz}, {Lattanzi}, {van Leeuwen}, {Bakker}, {Cacciari},
  {Casta{\~n}eda}, {De Angeli}, {Ducourant}, {Fabricius}, {Fouesneau},
  {Fr{\'e}mat}, {Guerra}, {Guerrier}, {Guiraud}, {Jean-Antoine Piccolo},
  {Masana}, {Messineo}, {Mowlavi}, {Nicolas}, {Nienartowicz}, {Pailler},
  {Panuzzo}, {Riclet}, {Roux}, {Seabroke}, {Sordo}, {Tanga}, {Th{\'e}venin},
  {Gracia-Abril}, {Portell}, {Teyssier}, {Altmann}, {Andrae}, {Bellas-Velidis},
  {Benson}, {Berthier}, {Blomme}, {Brugaletta}, {Burgess}, {Busso}, {Carry},
  {Cellino}, {Cheek}, {Clementini}, {Damerdji}, {Davidson}, {Delchambre},
  {Dell'Oro}, {Fern{\'a}ndez-Hern{\'a}ndez}, {Galluccio}, {Garc{\'\i}a-Lario},
  {Garcia-Reinaldos}, {Gonz{\'a}lez-N{\'u}{\~n}ez}, {Gosset}, {Haigron},
  {Halbwachs}, {Hambly}, {Harrison}, {Hatzidimitriou}, {Heiter},
  {Hern{\'a}ndez}, {Hestroffer}, {Hodgkin}, {Holl}, {Jan{\ss}en}, {Jevardat de
  Fombelle}, {Jordan}, {Krone-Martins}, {Lanzafame}, {L{\"o}ffler}, {Lorca},
  {Manteiga}, {Marchal}, {Marrese}, {Moitinho}, {Mora}, {Muinonen}, {Osborne},
  {Pancino}, {Pauwels}, {Petit}, {Recio-Blanco}, {Richards}, {Riello},
  {Rimoldini}, {Robin}, {Roegiers}, {Rybizki}, {Sarro}, {Siopis}, {Smith},
  {Sozzetti}, {Ulla}, {Utrilla}, {van Leeuwen}, {van Reeven}, {Abbas}, {Abreu
  Aramburu}, {Accart}, {Aerts}, {Aguado}, {Ajaj}, {Altavilla}, {{\'A}lvarez},
  {{\'A}lvarez Cid-Fuentes}, {Alves}, {Anderson}, {Anglada Varela}, {Antoja},
  {Audard}, {Baines}, {Baker}, {Balaguer-N{\'u}{\~n}ez}, {Balbinot}, {Balog},
  {Barache}, {Barbato}, {Barros}, {Barstow}, {Bartolom{\'e}}, {Bassilana},
  {Bauchet}, {Baudesson-Stella}, {Becciani}, {Bellazzini}, {Bernet}, {Bertone},
  {Bianchi}, {Blanco-Cuaresma}, {Boch}, {Bombrun}, {Bossini}, {Bouquillon},
  {Bragaglia}, {Bramante}, {Breedt}, {Bressan}, {Brouillet}, {Bucciarelli},
  {Burlacu}, {Busonero}, {Butkevich}, {Buzzi}, {Caffau}, {Cancelliere},
  {C{\'a}novas}, {Cantat-Gaudin}, {Carballo}, {Carlucci}, {Carnerero},
  {Carrasco}, {Casamiquela}, {Castellani}, {Castro-Ginard}, {Castro Sampol},
  {Chaoul}, {Charlot}, {Chemin}, {Chiavassa}, {Cioni}, {Comoretto}, {Cooper},
  {Cornez}, {Cowell}, {Crifo}, {Crosta}, {Crowley}, {Dafonte}, {Dapergolas},
  {David}, {David}, {de Laverny}, {De Luise}, {De March}, {De Ridder}, {de
  Souza}, {de Teodoro}, {de Torres}, {del Peloso}, {del Pozo}, {Delbo},
  {Delgado}, {Delgado}, {Delisle}, {Di Matteo}, {Diakite}, {Diener},
  {Distefano}, {Dolding}, {Eappachen}, {Edvardsson}, {Enke}, {Esquej}, {Fabre},
  {Fabrizio}, {Faigler}, {Fedorets}, {Fernique}, {Fienga}, {Figueras},
  {Fouron}, {Fragkoudi}, {Fraile}, {Franke}, {Gai}, {Garabato},
  {Garcia-Gutierrez}, {Garc{\'\i}a-Torres}, {Garofalo}, {Gavras}, {Gerlach},
  {Geyer}, {Giacobbe}, {Gilmore}, {Girona}, {Giuffrida}, {Gomel}, {Gomez},
  {Gonzalez-Santamaria}, {Gonz{\'a}lez-Vidal}, {Granvik},
  {Guti{\'e}rrez-S{\'a}nchez}, {Guy}, {Hauser}, {Haywood}, {Helmi}, {Hidalgo},
  {Hilger}, {H{\l}adczuk}, {Hobbs}, {Holland}, {Huckle}, {Jasniewicz},
  {Jonker}, {Juaristi Campillo}, {Julbe}, {Karbevska}, {Kervella}, {Khanna},
  {Kochoska}, {Kontizas}, {Kordopatis}, {Korn}, {Kostrzewa-Rutkowska},
  {Kruszy{\'n}ska}, {Lambert}, {Lanza}, {Lasne}, {Le Campion}, {Le Fustec},
  {Lebreton}, {Lebzelter}, {Leccia}, {Leclerc}, {Lecoeur-Taibi}, {Liao},
  {Licata}, {Lindstr{\o}m}, {Lister}, {Livanou}, {Lobel}, {Madrero Pardo},
  {Managau}, {Mann}, {Marchant}, {Marconi}, {Marcos Santos}, {Marinoni},
  {Marocco}, {Marshall}, {Martin Polo}, {Mart{\'\i}n-Fleitas}, {Masip},
  {Massari}, {Mastrobuono-Battisti}, {Mazeh}, {McMillan}, {Messina},
  {Michalik}, {Millar}, {Mints}, {Molina}, {Molinaro}, {Moln{\'a}r},
  {Montegriffo}, {Mor}, {Morbidelli}, {Morel}, {Morris}, {Mulone}, {Munoz},
  {Muraveva}, {Murphy}, {Musella}, {Noval}, {Ord{\'e}novic}, {Orr{\`u}},
  {Osinde}, {Pagani}, {Pagano}, {Palaversa}, {Palicio}, {Panahi}, {Pawlak},
  {Pe{\~n}alosa Esteller}, {Penttil{\"a}}, {Piersimoni}, {Pineau}, {Plachy},
  {Plum}, {Poggio}, {Poretti}, {Poujoulet}, {Pr{\v{s}}a}, {Pulone}, {Racero},
  {Ragaini}, {Rainer}, {Raiteri}, {Rambaux}, {Ramos}, {Ramos-Lerate}, {Re
  Fiorentin}, {Regibo}, {Reyl{\'e}}, {Ripepi}, {Riva}, {Rixon}, {Robichon},
  {Robin}, {Roelens}, {Rohrbasser}, {Romero-G{\'o}mez}, {Rowell}, {Royer},
  {Rybicki}, {Sadowski}, {Sagrist{\`a} Sell{\'e}s}, {Sahlmann}, {Salgado},
  {Salguero}, {Samaras}, {Sanchez Gimenez}, {Sanna}, {Santove{\~n}a},
  {Sarasso}, {Schultheis}, {Sciacca}, {Segol}, {Segovia}, {S{\'e}gransan},
  {Semeux}, {Shahaf}, {Siddiqui}, {Siebert}, {Siltala}, {Slezak}, {Smart},
  {Solano}, {Solitro}, {Souami}, {Souchay}, {Spagna}, {Spoto}, {Steele},
  {Steidelm{\"u}ller}, {Stephenson}, {S{\"u}veges}, {Szabados}, {Szegedi-Elek},
  {Taris}, {Tauran}, {Taylor}, {Teixeira}, {Thuillot}, {Tonello}, {Torra},
  {Torra}, {Turon}, {Unger}, {Vaillant}, {van Dillen}, {Vanel}, {Vecchiato},
  {Viala}, {Vicente}, {Voutsinas}, {Weiler}, {Wevers}, {Wyrzykowski}, {Yoldas},
  {Yvard}, {Zhao}, {Zorec}, {Zucker}, {Zurbach}, \&
  {Zwitter}}]{2021A&A...649A...1G}
{Gaia Collaboration}, {Brown}, A.~G.~A., {Vallenari}, A., {et~al.} 2021, \aap,
  649, A1, \dodoi{10.1051/0004-6361/202039657}

\bibitem[{{Hale}(1994)}]{1994AJ....107..306H}
{Hale}, A. 1994, \aj, 107, 306, \dodoi{10.1086/116855}

\bibitem[{{Hall} {et~al.}(2021){Hall}, {Davies}, {van Saders}, {Nielsen},
  {Lund}, {Chaplin}, {Garc{\'\i}a}, {Amard}, {Breimann}, {Khan}, {See}, \&
  {Tayar}}]{2021NatAs...5..707H}
{Hall}, O.~J., {Davies}, G.~R., {van Saders}, J., {et~al.} 2021, Nature
  Astronomy, 5, 707, \dodoi{10.1038/s41550-021-01335-x}

\bibitem[{{Hatzes} {et~al.}(2003){Hatzes}, {Cochran}, {Endl}, {McArthur},
  {Paulson}, {Walker}, {Campbell}, \& {Yang}}]{2003ApJ...599.1383H}
{Hatzes}, A.~P., {Cochran}, W.~D., {Endl}, M., {et~al.} 2003, \apj, 599, 1383,
  \dodoi{10.1086/379281}

\bibitem[{{Hayashi}(1981)}]{1981PThPS..70...35H}
{Hayashi}, C. 1981, Progress of Theoretical Physics Supplement, 70, 35,
  \dodoi{10.1143/PTPS.70.35}

\bibitem[{{Howard} {et~al.}(2010){Howard}, {Johnson}, {Marcy}, {Fischer},
  {Wright}, {Bernat}, {Henry}, {Peek}, {Isaacson}, {Apps}, {Endl}, {Cochran},
  {Valenti}, {Anderson}, \& {Piskunov}}]{2010ApJ...721.1467H}
{Howard}, A.~W., {Johnson}, J.~A., {Marcy}, G.~W., {et~al.} 2010, \apj, 721,
  1467, \dodoi{10.1088/0004-637X/721/2/1467}

\bibitem[{Hunter(2007)}]{Hunter:2007}
Hunter, J.~D. 2007, Computing in Science \& Engineering, 9, 90,
  \dodoi{10.1109/MCSE.2007.55}

\bibitem[{{Johnson} {et~al.}(2017){Johnson}, {Petigura}, {Fulton}, {Marcy},
  {Howard}, {Isaacson}, {Hebb}, {Cargile}, {Morton}, {Weiss}, {Winn}, {Rogers},
  {Sinukoff}, \& {Hirsch}}]{2017AJ....154..108J}
{Johnson}, J.~A., {Petigura}, E.~A., {Fulton}, B.~J., {et~al.} 2017, \aj, 154,
  108, \dodoi{10.3847/1538-3881/aa80e7}

\bibitem[{Jones {et~al.}(2001)Jones, Oliphant, Peterson, {et~al.}}]{scipy}
Jones, E., Oliphant, T., Peterson, P., {et~al.} 2001, {SciPy}: Open source
  scientific tools for {Python}.
\newblock \url{http://www.scipy.org/}

\bibitem[{{Justesen} \& {Albrecht}(2020)}]{2020A&A...642A.212J}
{Justesen}, A.~B., \& {Albrecht}, S. 2020, \aap, 642, A212,
  \dodoi{10.1051/0004-6361/202039138}

\bibitem[{{Konopacky} {et~al.}(2016){Konopacky}, {Marois}, {Macintosh},
  {Galicher}, {Barman}, {Metchev}, \& {Zuckerman}}]{2016AJ....152...28K}
{Konopacky}, Q.~M., {Marois}, C., {Macintosh}, B.~A., {et~al.} 2016, \aj, 152,
  28, \dodoi{10.3847/0004-6256/152/2/28}

\bibitem[{{Kozai}(1962)}]{1962AJ.....67..591K}
{Kozai}, Y. 1962, \aj, 67, 591, \dodoi{10.1086/108790}

\bibitem[{{Kratter} \& {Lodato}(2016)}]{2016ARA&A..54..271K}
{Kratter}, K., \& {Lodato}, G. 2016, \araa, 54, 271,
  \dodoi{10.1146/annurev-astro-081915-023307}

\bibitem[{{Kraus} {et~al.}(2016){Kraus}, {Ireland}, {Huber}, {Mann}, \&
  {Dupuy}}]{2016AJ....152....8K}
{Kraus}, A.~L., {Ireland}, M.~J., {Huber}, D., {Mann}, A.~W., \& {Dupuy}, T.~J.
  2016, \aj, 152, 8, \dodoi{10.3847/0004-6256/152/1/8}

\bibitem[{{Kuzuhara} {et~al.}(2022){Kuzuhara}, {Currie}, {Takarada}, {Brandt},
  {Sato}, {Uyama}, {Janson}, {Chilcote}, {Tobin}, {Lawson}, {Hori}, {Guyon},
  {Groff}, {Lozi}, {Vievard}, {Sahoo}, {Deo}, {Jovanovic}, {Ahn}, {Martinache},
  {Skaf}, {Akiyama}, {Norris}, {Bonnefoy}, {He{\l}miniak}, {Kudo}, {McElwain},
  {Samland}, {Wagner}, {Wisniewski}, {Knapp}, {Kwon}, {Nishikawa}, {Serabyn},
  {Hayashi}, \& {Tamura}}]{2022arXiv220502729K}
{Kuzuhara}, M., {Currie}, T., {Takarada}, T., {et~al.} 2022, arXiv e-prints,
  arXiv:2205.02729.
\newblock \doarXiv{2205.02729}

\bibitem[{{Lee} {et~al.}(2014){Lee}, {Chiang}, \&
  {Ormel}}]{2014ApJ...797...95L}
{Lee}, E.~J., {Chiang}, E., \& {Ormel}, C.~W. 2014, \apj, 797, 95,
  \dodoi{10.1088/0004-637X/797/2/95}

\bibitem[{{Li} {et~al.}(2021){Li}, {Brandt}, {Brandt}, {Dupuy}, {Michalik},
  {Jensen-Clem}, {Zeng}, {Faherty}, \& {Mitra}}]{2021AJ....162..266L}
{Li}, Y., {Brandt}, T.~D., {Brandt}, G.~M., {et~al.} 2021, \aj, 162, 266,
  \dodoi{10.3847/1538-3881/ac27ab}

\bibitem[{{Lidov}(1962)}]{1962P&SS....9..719L}
{Lidov}, M.~L. 1962, \planss, 9, 719, \dodoi{10.1016/0032-0633(62)90129-0}

\bibitem[{{Lissauer}(1987)}]{1987Icar...69..249L}
{Lissauer}, J.~J. 1987, \icarus, 69, 249, \dodoi{10.1016/0019-1035(87)90104-7}

\bibitem[{{Lissauer} {et~al.}(2011){Lissauer}, {Ragozzine}, {Fabrycky},
  {Steffen}, {Ford}, {Jenkins}, {Shporer}, {Holman}, {Rowe}, {Quintana},
  {Batalha}, {Borucki}, {Bryson}, {Caldwell}, {Carter}, {Ciardi}, {Dunham},
  {Fortney}, {Gautier}, {Howell}, {Koch}, {Latham}, {Marcy}, {Morehead}, \&
  {Sasselov}}]{2011ApJS..197....8L}
{Lissauer}, J.~J., {Ragozzine}, D., {Fabrycky}, D.~C., {et~al.} 2011, \apjs,
  197, 8, \dodoi{10.1088/0067-0049/197/1/8}

\bibitem[{{Lubow} {et~al.}(2015){Lubow}, {Martin}, \&
  {Nixon}}]{2015ApJ...800...96L}
{Lubow}, S.~H., {Martin}, R.~G., \& {Nixon}, C. 2015, \apj, 800, 96,
  \dodoi{10.1088/0004-637X/800/2/96}

\bibitem[{{Mack} {et~al.}(2018){Mack}, {Strassmeier}, {Ilyin}, {Schuler},
  {Spada}, \& {Barnes}}]{2018A&A...612A..46M}
{Mack}, C.~E., {Strassmeier}, K.~G., {Ilyin}, I., {et~al.} 2018, \aap, 612,
  A46, \dodoi{10.1051/0004-6361/201731634}

\bibitem[{{Manara} {et~al.}(2019){Manara}, {Tazzari}, {Long}, {Herczeg},
  {Lodato}, {Rota}, {Cazzoletti}, {van der Plas}, {Pinilla}, {Dipierro},
  {Edwards}, {Harsono}, {Johnstone}, {Liu}, {Menard}, {Nisini}, {Ragusa},
  {Boehler}, \& {Cabrit}}]{2019A&A...628A..95M}
{Manara}, C.~F., {Tazzari}, M., {Long}, F., {et~al.} 2019, \aap, 628, A95,
  \dodoi{10.1051/0004-6361/201935964}

\bibitem[{{Markwardt}(2009)}]{2009ASPC..411..251M}
{Markwardt}, C.~B. 2009, in Astronomical Society of the Pacific Conference
  Series, Vol. 411, Astronomical Data Analysis Software and Systems XVIII, ed.
  D.~A. {Bohlender}, D.~{Durand}, \& P.~{Dowler}, 251

\bibitem[{{Mazeh} {et~al.}(2015){Mazeh}, {Perets}, {McQuillan}, \&
  {Goldstein}}]{2015ApJ...801....3M}
{Mazeh}, T., {Perets}, H.~B., {McQuillan}, A., \& {Goldstein}, E.~S. 2015,
  \apj, 801, 3, \dodoi{10.1088/0004-637X/801/1/3}

\bibitem[{{Mills} \& {Fabrycky}(2017)}]{2017ApJ...838L..11M}
{Mills}, S.~M., \& {Fabrycky}, D.~C. 2017, \apjl, 838, L11,
  \dodoi{10.3847/2041-8213/aa6543}

\bibitem[{{Miranda} \& {Lai}(2015)}]{2015MNRAS.452.2396M}
{Miranda}, R., \& {Lai}, D. 2015, \mnras, 452, 2396,
  \dodoi{10.1093/mnras/stv1450}

\bibitem[{{Moe} \& {Kratter}(2021)}]{2021MNRAS.507.3593M}
{Moe}, M., \& {Kratter}, K.~M. 2021, \mnras, 507, 3593,
  \dodoi{10.1093/mnras/stab2328}

\bibitem[{{Naoz} {et~al.}(2013){Naoz}, {Farr}, {Lithwick}, {Rasio}, \&
  {Teyssandier}}]{2013MNRAS.431.2155N}
{Naoz}, S., {Farr}, W.~M., {Lithwick}, Y., {Rasio}, F.~A., \& {Teyssandier}, J.
  2013, \mnras, 431, 2155, \dodoi{10.1093/mnras/stt302}

\bibitem[{{Offner} {et~al.}(2022){Offner}, {Moe}, {Kratter}, {Sadavoy},
  {Jensen}, \& {Tobin}}]{2022arXiv220310066O}
{Offner}, S. S.~R., {Moe}, M., {Kratter}, K.~M., {et~al.} 2022, arXiv e-prints,
  arXiv:2203.10066.
\newblock \doarXiv{2203.10066}

\bibitem[{Oliphant(2006)}]{numpy}
Oliphant, T. 2006, {NumPy}: A guide to {NumPy}, USA: Trelgol Publishing.
\newblock \url{http://www.numpy.org/}

\bibitem[{{Pearce} {et~al.}(2020){Pearce}, {Kraus}, {Dupuy}, {Mann}, {Newton},
  {Tofflemire}, \& {Vanderburg}}]{2020ApJ...894..115P}
{Pearce}, L.~A., {Kraus}, A.~L., {Dupuy}, T.~J., {et~al.} 2020, \apj, 894, 115,
  \dodoi{10.3847/1538-4357/ab8389}

\bibitem[{P\'erez \& Granger(2007)}]{PER-GRA:2007}
P\'erez, F., \& Granger, B.~E. 2007, Computing in Science and Engineering, 9,
  21, \dodoi{10.1109/MCSE.2007.53}

\bibitem[{{Rafikov} \& {Silsbee}(2015)}]{2015ApJ...798...69R}
{Rafikov}, R.~R., \& {Silsbee}, K. 2015, \apj, 798, 69,
  \dodoi{10.1088/0004-637X/798/2/69}

\bibitem[{{Raghavan} {et~al.}(2010){Raghavan}, {McAlister}, {Henry}, {Latham},
  {Marcy}, {Mason}, {Gies}, {White}, \& {ten Brummelaar}}]{2010ApJS..190....1R}
{Raghavan}, D., {McAlister}, H.~A., {Henry}, T.~J., {et~al.} 2010, \apjs, 190,
  1, \dodoi{10.1088/0067-0049/190/1/1}

\bibitem[{{Ricker} {et~al.}(2015){Ricker}, {Winn}, {Vanderspek}, {Latham},
  {Bakos}, {Bean}, {Berta-Thompson}, {Brown}, {Buchhave}, {Butler}, {Butler},
  {Chaplin}, {Charbonneau}, {Christensen-Dalsgaard}, {Clampin}, {Deming},
  {Doty}, {De Lee}, {Dressing}, {Dunham}, {Endl}, {Fressin}, {Ge}, {Henning},
  {Holman}, {Howard}, {Ida}, {Jenkins}, {Jernigan}, {Johnson}, {Kaltenegger},
  {Kawai}, {Kjeldsen}, {Laughlin}, {Levine}, {Lin}, {Lissauer}, {MacQueen},
  {Marcy}, {McCullough}, {Morton}, {Narita}, {Paegert}, {Palle}, {Pepe},
  {Pepper}, {Quirrenbach}, {Rinehart}, {Sasselov}, {Sato}, {Seager},
  {Sozzetti}, {Stassun}, {Sullivan}, {Szentgyorgyi}, {Torres}, {Udry}, \&
  {Villasenor}}]{2015JATIS...1a4003R}
{Ricker}, G.~R., {Winn}, J.~N., {Vanderspek}, R., {et~al.} 2015, Journal of
  Astronomical Telescopes, Instruments, and Systems, 1, 014003,
  \dodoi{10.1117/1.JATIS.1.1.014003}

\bibitem[{{Service} {et~al.}(2016){Service}, {Lu}, {Campbell}, {Sitarski},
  {Ghez}, \& {Anderson}}]{2016PASP..128i5004S}
{Service}, M., {Lu}, J.~R., {Campbell}, R., {et~al.} 2016, \pasp, 128, 095004,
  \dodoi{10.1088/1538-3873/128/967/095004}

\bibitem[{{Shetrone} {et~al.}(2007){Shetrone}, {Cornell}, {Fowler}, {Gaffney},
  {Laws}, {Mader}, {Mason}, {Odewahn}, {Roman}, {Rostopchin}, {Schneider},
  {Umbarger}, \& {Westfall}}]{2007PASP..119..556S}
{Shetrone}, M., {Cornell}, M.~E., {Fowler}, J.~R., {et~al.} 2007, \pasp, 119,
  556, \dodoi{10.1086/519291}

\bibitem[{{Sigalotti} {et~al.}(2018){Sigalotti}, {Cruz}, {Gabbasov}, {Klapp},
  \& {Ram{\'\i}rez-Velasquez}}]{2018ApJ...857...40S}
{Sigalotti}, L. D.~G., {Cruz}, F., {Gabbasov}, R., {Klapp}, J., \&
  {Ram{\'\i}rez-Velasquez}, J. 2018, \apj, 857, 40,
  \dodoi{10.3847/1538-4357/aab619}

\bibitem[{{Silsbee} \& {Rafikov}(2015)}]{2015ApJ...798...71S}
{Silsbee}, K., \& {Rafikov}, R.~R. 2015, \apj, 798, 71,
  \dodoi{10.1088/0004-637X/798/2/71}

\bibitem[{{Silva Aguirre} {et~al.}(2015){Silva Aguirre}, {Davies}, {Basu},
  {Christensen-Dalsgaard}, {Creevey}, {Metcalfe}, {Bedding}, {Casagrande},
  {Handberg}, {Lund}, {Nissen}, {Chaplin}, {Huber}, {Serenelli}, {Stello}, {Van
  Eylen}, {Campante}, {Elsworth}, {Gilliland}, {Hekker}, {Karoff}, {Kawaler},
  {Kjeldsen}, \& {Lundkvist}}]{2015MNRAS.452.2127S}
{Silva Aguirre}, V., {Davies}, G.~R., {Basu}, S., {et~al.} 2015, \mnras, 452,
  2127, \dodoi{10.1093/mnras/stv1388}

\bibitem[{{Sozzetti} {et~al.}(2009){Sozzetti}, {Torres}, {Latham}, {Stefanik},
  {Korzennik}, {Boss}, {Carney}, \& {Laird}}]{2009ApJ...697..544S}
{Sozzetti}, A., {Torres}, G., {Latham}, D.~W., {et~al.} 2009, \apj, 697, 544,
  \dodoi{10.1088/0004-637X/697/1/544}

\bibitem[{{Tayar} {et~al.}(2022){Tayar}, {Claytor}, {Huber}, \& {van
  Saders}}]{2022ApJ...927...31T}
{Tayar}, J., {Claytor}, Z.~R., {Huber}, D., \& {van Saders}, J. 2022, \apj,
  927, 31, \dodoi{10.3847/1538-4357/ac4bbc}

\bibitem[{{Th{\'e}bault} {et~al.}(2006){Th{\'e}bault}, {Marzari}, \&
  {Scholl}}]{2006Icar..183..193T}
{Th{\'e}bault}, P., {Marzari}, F., \& {Scholl}, H. 2006, \icarus, 183, 193,
  \dodoi{10.1016/j.icarus.2006.01.022}

\bibitem[{{Tobin} {et~al.}(2016){Tobin}, {Kratter}, {Persson}, {Looney},
  {Dunham}, {Segura-Cox}, {Li}, {Chandler}, {Sadavoy}, {Harris}, {Melis}, \&
  {P{\'e}rez}}]{2016Natur.538..483T}
{Tobin}, J.~J., {Kratter}, K.~M., {Persson}, M.~V., {et~al.} 2016, \nat, 538,
  483, \dodoi{10.1038/nature20094}

\bibitem[{{Tokovinin}(2018)}]{2018AJ....155..160T}
{Tokovinin}, A. 2018, \aj, 155, 160, \dodoi{10.3847/1538-3881/aab102}

\bibitem[{{Tull}(1998)}]{1998SPIE.3355..387T}
{Tull}, R.~G. 1998, in Society of Photo-Optical Instrumentation Engineers
  (SPIE) Conference Series, Vol. 3355, Optical Astronomical Instrumentation,
  ed. S.~{D'Odorico}, 387--398

\bibitem[{{Vousden} {et~al.}(2016){Vousden}, {Farr}, \&
  {Mandel}}]{2016MNRAS.455.1919V}
{Vousden}, W.~D., {Farr}, W.~M., \& {Mandel}, I. 2016, \mnras, 455, 1919,
  \dodoi{10.1093/mnras/stv2422}

\bibitem[{{Wang} {et~al.}(2014){Wang}, {Xie}, {Barclay}, \&
  {Fischer}}]{2014ApJ...783....4W}
{Wang}, J., {Xie}, J.-W., {Barclay}, T., \& {Fischer}, D.~A. 2014, \apj, 783,
  4, \dodoi{10.1088/0004-637X/783/1/4}

\bibitem[{{Weidenschilling}(1977)}]{1977Ap&SS..51..153W}
{Weidenschilling}, S.~J. 1977, \apss, 51, 153, \dodoi{10.1007/BF00642464}

\bibitem[{{Wilson}(1941)}]{1941ApJ....93...29W}
{Wilson}, O.~C. 1941, \apj, 93, 29, \dodoi{10.1086/144239}

\bibitem[{{Wizinowich}(2013)}]{2013PASP..125..798W}
{Wizinowich}, P. 2013, \pasp, 125, 798, \dodoi{10.1086/671425}

\bibitem[{{Yelda} {et~al.}(2010){Yelda}, {Lu}, {Ghez}, {Clarkson}, {Anderson},
  {Do}, \& {Matthews}}]{2010ApJ...725..331Y}
{Yelda}, S., {Lu}, J.~R., {Ghez}, A.~M., {et~al.} 2010, \apj, 725, 331,
  \dodoi{10.1088/0004-637X/725/1/331}

\bibitem[{{Zanazzi} \& {Lai}(2018)}]{2018MNRAS.477.5207Z}
{Zanazzi}, J.~J., \& {Lai}, D. 2018, \mnras, 477, 5207,
  \dodoi{10.1093/mnras/sty951}

\bibitem[{{Zeng} {et~al.}(2021){Zeng}, {Brandt}, {Li}, {Dupuy}, {Li}, {Brandt},
  {Farihi}, {Horner}, {Wittenmyer}, {Paul. Butler}, {Tinney}, {Carter},
  {Wright}, {Jones}, \& {O'Toole}}]{2021arXiv211206394Z}
{Zeng}, Y., {Brandt}, T.~D., {Li}, G., {et~al.} 2021, arXiv e-prints,
  arXiv:2112.06394.
\newblock \doarXiv{2112.06394}

\bibitem[{{Ziegler} {et~al.}(2021){Ziegler}, {Tokovinin}, {Latiolais},
  {Brice{\~n}o}, {Law}, \& {Mann}}]{2021AJ....162..192Z}
{Ziegler}, C., {Tokovinin}, A., {Latiolais}, M., {et~al.} 2021, \aj, 162, 192,
  \dodoi{10.3847/1538-3881/ac17f6}

\end{thebibliography}

\clearpage

\clearpage
\tabletypesize{\scriptsize}

\end{document}